\documentclass[10pt]{iopart}
\usepackage{iopams,epsf,epsfig}
\newcommand{\gam}           {\gamma}
\newcommand{\kap}           {\kappa}
\newcommand{\wa}    {\omega_{a}}
\newcommand{\wc}    {\omega_{c}}
\newcommand{\wat}    {\widetilde{{\omega}}_{a}}
\newcommand{\wct}    {\widetilde{{\omega}}_{c}}
\newcommand{\alpho}           {\alpha_0}
\newcommand{\beto}           {\beta_{0}}
\newcommand{\GamA}           {\Gamma_a}
\newcommand{\KC}           {K_c}
\newcommand{\WA}    {\Omega_{a}}
\newcommand{\WC}    {\Omega_{c}}
\newcommand{\WAt}    {\widetilde{{\Omega}}_{a}}

\newcommand{\Fcal}  {\mathcal {F}_{dip}}
\newcommand{\Liou}  {\mathcal {L}}
\newcommand{\Lsig}  {L_{\sigma}}
\newcommand{\LSig}  {L_{\Sigma}}
\newcommand{\La}  {L_{a}}
\newcommand{\Lb}  {L_{b}}
\newcommand{\Lc}  {L_{c}}
\newcommand{\keto}        {\left|0\right>}
\newcommand{\brao}        {\left<0\right|}
\newcommand{\aop}        {a}
\newcommand{\caop}       {a^{\dag}}
\newcommand{\cop}        {c}
\newcommand{\ccop}       {c^{\dag}}
\newcommand{\csig}       {\sigma^{\dag}}
\newcommand{\sig}       {\sigma}
\newcommand{\cSig}       {\Sigma^{\dag}}
\newcommand{\Sig}       {\Sigma}
\newcommand{\gdg}   {g\partial_g}
\newcommand{\dg}   {\partial_g}

\newcommand{\beq}   {\begin{equation}}
\newcommand{\eeq}   {\end{equation}}
\newcommand{\beqz}  {\setlength{\mathindent}{0cm}\begin{equation}}
\newcommand{\eeqz}  {\end{equation}}
\newcommand{\ber}   {\begin{eqnarray}}
\newcommand{\eer}   {\end{eqnarray}}
\newcommand{\bers}  {\begin{eqnarray*}}
\newcommand{\eers}  {\end{eqnarray*}}

\begin{document}
\title[Driven cavities in the strong coupling regime: Structure and
dynamics]{On the suppression of the diffusion
and the quantum nature of a cavity mode. Optical bistability;
forces and friction in driven cavities}
\author{Karim Murr\footnote[3]{e-mail:
karim.elmurr@unicam.it} }

\address{INFM, Dipartimento di Fisica, Universit\`a di Camerino,
I-62032 Camerino, Italy}

\begin{abstract}
A new analytical method is presented here, offering a physical view of
driven cavities where the external field cannot be neglected. We
introduce a new dimensionless complex parameter, intrinsically
linked to the cooperativity parameter of optical bistability, and
analogous to the scaled Rabbi frequency for driven systems where the
field is classical. Classes of steady states are iteratively constructed
and expressions for the diffusion and friction coefficients at lowest
order also derived. They have in most cases the same mathematical form
as their free-space analog. The method offers a semiclassical
explanation for two recent experiments of one atom trapping in a high
$\mathcal{Q}$ cavity where the excited state is significantly saturated.
Our results refute both claims of atom trapping by a quantized cavity
mode, single or not. Finally, it is argued that the parameter newly
constructed, as well as the groundwork of this method, are at
least companions of the cooperativity parameter and its mother
theory. In particular, we lay the stress on the apparently more
fundamental role of our structure parameter.
\end{abstract}

\pacs{32.80.Lg, 42.65.Pc, 31.15.Gy, 42.50.Fx}

\submitto{\jpb}


\section{Introduction}
In 1994, Kimble (1994) quoted the necessity of theoretical work to
be done in order to understand the dynamics of an atom in a cavity
where the mode is not a prescribed quantity but rather a fully
playing actor. A forefront difficulty is to draw an efficient
theoretical framework that treats the strong coupling regime in
open systems, externally fed, and dissipating by cavity decay
$\kap$ and/or spontaneous emission $\gam$. The first area of
cavity QED is usually devoted to coherent aspects of the coupling
$g$ between the atom and the cavity mode. It is well treated
within the dressed states formalism (Cohen-Tannoudji 1990), a
vision that contains in block all the information about the
quantized field plus the atom and where, at first, dissipation and
driving are not accounted for. On the other hand, driving a cavity
leads to dissipative structures. Those are well understood within
the semiclassical theory of optical bistability (Lugiato and
Narducci 1990). Semiclassical means here that the field is
factorized from the internal state of the atom, in the operatorial
sense. This corresponds in bistability theory to treat the field
classically by means of Maxwell-Bloch equations. The smallness of
the decay rates ($\kap,\gam$) allows the field to be kept inside
the resonator boundaries, and hence, a long interactivity occurs
before the escape of one photon or one atom. A difficulty arises
as there is then a competition between the scaled Rabbi frequency
of the dressed states (coherent part) and the Bonifacio-Lugiato
cooperativity parameter $C=g^2\big/2\kap\gam$ (Bonifacio and
Lugiato 1978,1975).

 Recently, two major experiments reported the trapping of an atom
in a cavity containing about one photon on average (Hood \etal 2000,
Pinkse \etal 2000). In both papers, the trapping times of about a
millisecond and the registered signals were considered to be a
signature of the quantum nature of the field. Another study (Doherty
\etal 2000) suggested by theoretical and numerical arguments that the
experiment of (Pinkse \etal 2000) is well understood without
quantization assumption of the field. One argument was to compare the
diffusion coefficient and the trapping potential which they found to be
close to the free space ones (Gordon and Ashkin 1980). In contrast, the
experiment of (Hood \etal 2000) deals with a parameter regime where the
diffusion is suppressed by nearly a factor of ten when compared to the
free space case. (Hood \etal 2000, Doherty \etal 2000) conclude that
such high suppression is a signature of the dynamics embedded in the
Jaynes-Cummings ladder of dressed states, pointing out indirectly on the
quantum nature of the trapping field. (Hood \etal 2000) support their
conclusions by citing other experiments, including (Hood \etal 1998),
that found good agreement between the full quantum calculus and the
experimental data of a heterodyne signal, on the single atom level,
whereas the semiclassical bistability state equation fails.

 In parallel, (Pinkse \etal 2000) attribute the trapping mechanisms to
cavity-induced cooling (Horak \etal 1997, Hechenblaikner \etal 1998).
That mechanism is based on the weak-field limit and is extendable
(Domokos \etal 2001) to higher photon number provided a low saturation
is kept. Their work follows and precedes several studies concerned with
the mechanical effects of light on atoms within a cavity (see e.g
Doherty \etal 1997, Vuleti\'c and Chu 2000). A better understanding
together with an extension of the analytical calculus to higher
intensities and/or saturation, on the single atom level, seems difficult
as one expects, either the inclusion of multiphoton processes within the
dressed states formalism, or an extension of the bistability state
equation.

 Having drawn the different areas to be studied, we bear in mind that
the need for the full quantum calculation to fit some experimental data
does not necessarily mean that one actually deals with quantum
 mechanics,
rather, it indicates that the known semiclassical theories are
insufficient. We demonstrate in this paper that both experiments of
(Pinkse \etal 2000, Hood \etal 2000) are substantially explained by a
semiclassical approach. A strong result is that, \emph{even if one tunes
the external laser resonantly to a dressed state, neither experiment
needs the ladder of dressed states to be explained}. That is true for
elementary steady state quantities as well as for the diffusion and
friction coefficients. To that purpose, we develop a fully analytical
framework that allows one to construct the needed quantities, structure
and dynamics, by taking increasing orders of the coupling $g$, without
any dressed states consideration. We introduce a new physical parameter,
until now un-noticed, which is efficient for the treatment of driven
dissipative systems, and by essence is a cooperative quantity. The
method, in this paper, is limited to the explicit construction of
factorized states that are an iterative version of the bistability state
equation. That is possible by introducing the idea of \emph{the referred
states}. The referred states allow the construction of the steady
states. Six steady states are derived, their common structure is given
in section \ref{background}. Those are then grouped into two distinct
families, \emph{the bounced states} \sref{bounce}, and \emph{the
polarized states} \sref{polarized}, typically valid for increasing
saturation and photon number. Our parameter is introduced in subsection
\ref{introNU} and is discussed in subsection \ref{ping-pong}.
\Sref{sectionOB} draws mathematical remarks on the role of our parameter
in bistability theory. Finally, we derive three expressions for the
diffusion in \sref{diffusion} and two expressions for the friction in
\sref{friction}. Diffusion and friction are derived in such a way as to
be independent of any explicit expression of the states.
\subsection{The physical system}
\label{position}
In dealing with cavity QED one usually uses the driven Jaynes-Cummings
Hamiltonian for a 2 states atom represented by the lowering and rising
operators ($\sig,\csig$), $\sig\!=\!|g\rangle\langle e|$, where
$|g\rangle$ and $|e\rangle$ stand for the ground and excited states
respectively, and a single mode with annihilation and creation operators
($\aop,\caop$). We consider the so-called rotating wave approximation
and focus only on the internal state of the atom interacting with one
cavity mode. The master equation for the reduced density matrix $\rho\,$
of the system is governed by the Liouvillian $\Liou\ (\hslash=1)$:
\numparts
\beq
\label{Liou}
\dot\rho=\Liou \rho ,\quad \Liou\rho = -i[ H,\rho
]+\gam\Lsig\rho + \kappa\La\rho\ ,
\eeq
with the dissipative parts written in the form
$\,\,\displaystyle{\Lb\rho= 2b\rho b^\dag-\rho b^\dag b-b^\dag
b\rho},\,\,$ for any operator b, and with the Hamiltonian (in the frame
rotating with the probe frequency $\omega_{probe}$):
\beq
\label{Ham}
H = \wa\csig\sig + \wc\caop\aop + g(\caop\sig + \aop\csig) + E\caop +
E^*\aop\ .
\eeq
Here $\wa\,$ and $\wc\,$ are the atomic and cavity frequencies
respectively, detuned with respect to the probe frequency
$\omega_{probe}$. Although these detunings are usually labelled by
($\Delta_{a}, \Delta_{c}$), we keep our notations in order to have the
formulas more compact:
($\wa\!\!\equiv\!\omega_{atom}\!-\!\omega_{probe},\,
\wc\!\equiv\!\omega_{cav}\!-\!\omega_{probe}$). $\gam$ and $\kap$
represent respectively the atomic and cavity-mode decay rates, $E$ is
the strength of the probe field that coherently drives the cavity mode,
and $g$ is the coupling that we assume real and dependent on the
position of the center of mass of the atom. For illustration purposes we
assume a cosine dependence along the cavity axis $x$,
$g(x)\!=\!g_0\!\cos(2\pi x/\lambda)$, with wavelength $\lambda$ and
$g_0$ is the maximum of $g$. All the results presented are independent
of any particular form of the coupling $g$. Finally, the dipole force
operator $\Fcal$ reads:
\beq
\label{force}
\Fcal=-(\nabla g) F_d\,,\quad
F_d=\aop\csig+\caop\sig\ .
\eeq
\endnumparts
\subsection{Relevant dimensionless physical parameters}
\label{introNU}
In order to reach a consistent understanding of the physics of the
system, we look for the relevant parameters to be formed from
\eref{Liou}\eref{Ham}. For $E\!=\!0$, a natural description is provided
by the ladder of dressed states, whose structure is basically understood
by the square of the scaled Rabbi frequency
$\Omega_n^2=g^2(n+1)/\delta^2\equiv C_n$ (Cohen-Tannoudji 1990), where
$\delta=\wa-\wc$ is the detuning mismatch and $n$ the mode quantum
number. $C_n$ is independent of the probe frequency $\omega_{probe}$ and
built by atomic and mode parameters, where the latter has a discrete
quantum nature. If one includes a driving, the steady state mean photon
number in the cavity without the atom is given by the complex amplitude
$\alpho$:
\beq
\label{N0}
N_0\equiv\,|\alpho|^2\,,\quad\alpho=E/\wct\,,\qquad(\wct=\wc-i\kap)\
.
\eeq
The parameter $\alpho$ depends upon the probe through the strength $E$
and the cavity detuning $\wc$. When the driving is sufficiently high
$E\lesssim g_0$, a given dressed state becomes coupled to upper and
lower manifolds. It becomes hard to even deduce the mean photon number
$N_0$ from this formalism. Indeed, $\alpho$ is a continuous description
of the cavity mode that tends to dismantle (dilute) the discrete dressed
states by connecting them. When the atom is driven by a classical field
$-\alpha$ (e.g $-\alpho$), with a strength $-g\alpha$, the relevant
quantity is $g/\wat$ ($\wat=\wa-i\gam$) where $|\wat|^2/g^2$ gives the
amount of (normalized) field intensity needed to saturate the atom.
Symmetrically, when the atom is described by a classical dipole
oscillator with polarization $\sig\rightarrow\beta$, of strength
$g\beta$, a relevant parameter for the cavity mode is $g/\wct$,
whose square modulus represents the amount of field intensity that
is exchanged between the cavity and the atom (compare with
$\alpho$ \eref{N0}).

 The parameters ($E/\wct,g/\wat,g/\wct$) depend on the probe frequency
$\omega_{probe}$ and, contrary to $C_n$, they do not describe any
settled structure of the coupled-atom-mode system. A parameter
accounting for that is the cooperativity parameter $C=g^2\big/2\kap\gam$
in bistability theory, but this quantity does not include the detunings.
So we use the ratios $g/\wat$ and $g/\wct$ to introduce a new
dimensionless complex parameter:
\beq
\label{NU}
\nu\quad=\quad\displaystyle{g^2\over\wat\wct}\ .
\eeq
The parameter $\nu$ presents the same structure as the dressed states
parameter $C_n$ except that it depends on the probe frequency
$\omega_{probe}$ and contains no quantum signature of the cavity
mode. We shall stress throughout this paper that, when the cavity
mode is described classically, $\nu$ has a similar status as the
(squared) scaled Rabbi frequency $C_n$. As $\nu$ depends on the
probe frequency $\omega_{probe}$, then it is likely to be related
to the driving strength $E$. We show this below. Several
properties for $\nu$ should be mentioned: First it can be seen as
the ratio $(g/\wct)/(\wat/g)$, when squared it compares the number
of photons (average) provided by the atom to the number of photons
needed to saturate the atom. In dynamical terms, it measures the
ratio between the atomic dispersive shift $g^2/\wat$ and the field
complex frequency $\wct$. $\nu$ reduces to the cooperativity
parameter $-2C$ when the atom and cavity are at resonance with the
probe, and finally its imaginary part contains the mode-pulling
formula in laser theories $\Im(\nu)\propto\kap\wa+\gam\wc$
(Lugiato and Narducci 1990). Finally, the deep meaning for $\nu$ is
mathematically dealt with in subsection \ref{ping-pong}, it constitutes
in this paper a central physical parameter.
\subsection{A problem with the standard semiclassical factorization
procedure}
A description of the system that avoids the dressed states
formalism could be provided by the bistability state equation,
whose underlying theory is generally understood by invoking the
so-called semiclassical approximation. This approximation assumes
a factorization of the mean values of the product of atomic operators
$A$ and mode operators $C$. It is satisfied as far as the state of the
system can be factorized into a product $\rho=\rho_m\otimes\rho_a$,
where ($\rho_m,\rho_a$) are states for the mode (m) and the atom (a)
respectively. With the notations, $\tr(CA\rho)\equiv <\!CA\!>$,
$\tr(C\rho_m)\equiv<\!C\!>$, $\tr(A\rho_a)\equiv<\!A\!>$, that reads:
\beq
\label{fact}
<\!CA\!>=<\!C\!><\!A\!>\ .
\eeq
The Heisenberg equations of motion for
($<\!\dot a\!>,$$<\!\dot\sig\!>$$,<\!\dot\sig_z\!>$), with the
population difference $\displaystyle{\sig_z=\csig\sig\!-\!\sig\csig}$
($<\!\dot\csig\!>=<\!\dot\sig\!>^*$ is implicit), are written:
\numparts
\ber \label{obsig}
<\!\dot\sig\!>=-i\wat(<\!\sig\!>-\,\frac{g}{\wat}<\!\aop\sig_z\!>)\\
\label{oba} <\!\dot\aop\!>=-i\wct(<\!\aop\!>\,+\,\alpho\,+
\frac{g}{\wct}<\!\sig\!>)\\
\label{obsz}
<\!\dot\sig_z\!>=-2\gam -2\gam <\!\sig_z\!>
+2ig(<\!\caop\sig\!>\!-\!<\!\aop\csig\!>)\ .
\eer
\endnumparts
By the use of \eref{fact}, the closed system is solved at steady
state; by eliminating the atomic variables
($<\!\sig\!>,<\!\csig\!>,<\!\sig_z\!>$), one obtains the
\emph{bistability state equation} for the output field
$\alpha_{ob}\equiv-<\!a\!>$:
\ber
\label{ob}
\alpha_{ob}\,=\,\alpho\,[\,1-\frac{\nu}{1+s_{ob}}\,]^{-1} \,,\quad
s_{ob}=2\frac{g^2}{|\wat|^2}|\alpha_{ob}|^2\ ,
\eer
where $s_{ob}$ stands for the corresponding atomic saturation parameter,
$<\!\sig_z\!>=-1/(1\!+\!s_{ob})$. Notice that, written in terms of
$\nu$ \eref{NU}, the equation obtained is much clearer in form than
the usual expression (Lugiato and Narducci 1990, also see
\sref{sectionOB}). That equation is cubic in the steady state output
intensity $|\!\!<a>\!\!|^2=|\alpha_{ob}|^2$, thus it would possibly lead
to a hysteresis cycle. The non-linear S-shape of the bistability state
equation is up to now unobserved on the single atom level. (Hood \etal
1998) demonstrated a clear disagreement between the measured heterodyne
detection and the semiclassical bistability state equation, while the
data are still well explained by the exact steady state of the system
\eref{Liou}\eref{Ham}.
\section{General formalism}
\label{background}
\subsection{Iterative factorization procedure: Definition of the
referred states}
\label{Genesemi}
Our criticism to the above procedure is based on the remark that the
expectation values ($<\!\aop\!>,<\!\sig\!>,$$<\!\csig\!>,<\!\sig_z\!>$)
are \emph{all} considered as variables in
\eref{obsig}\eref{oba}\eref{obsz}. Whether they appear directly or they
are issued from the factorization assumption \eref{fact}. Mainly there
is no other choice than to deduce the explicit values by solving the
system to obtain equation \eref{ob}.

 Our approach is still to construct factorized states, but we treat
each dynamical equation \eref{obsig}\eref{oba}\eref{obsz}
independently and iteratively, with assumption \eref{fact} viewed
differently for each operator product of the form $CA$, at each
step. A specific example is provided to illustrate our method. A
low saturation regime is assumed, the exact steady state $\rho_s$
corresponds to an atom almost in the ground state $|g\rangle$,
$<\!\sig_z\!>_s\approx\!-1$ (vanishing saturation parameter).
Moreover, the exact mean values of product operators $CA$ can be
written in the form of a product of mean values. Eventually, the low
saturation regime is chosen such that one has for the bistability
state equation $s_{ob}\!\gg\!1$, hence \eref{ob} is a bad
(factorized state) approximation. Given these conditions, we
approximate the exact steady state $\rho_s$ by
$\rho_s\!\approx\!\rho_{m}\otimes|g\rangle\langle g|$, for the
mean value $<\!a\sig_z\!>$ \emph{only at this step},
$<\!a\sig_z\!>\approx\!-<\!a\!>$, to obtain $<\!\sig\!>$
\eref{obsig}, that is then replaced in \eref{oba} to obtain
$<\!a\!>=-\alpho/(1-\nu)$. We then assume that $<\!\caop\sig\!>$
factorizes according to \eref{fact}, to deduce $<\!\sig_z\!>$ from
\eref{obsz}. Those first steps give,
\numparts
\ber
\label{Ritsch}
\alpha=\frac{\alpho}{1-\nu}\ ,\quad <\!\sig_z\!>=-1+s\ ,\quad
<\!\sig\!>=\frac{g\alpha}{\wat}\ ,
\eer
where we have defined $\alpha\!\equiv\!-<\!a\!>$, and
$s=2g^2/|\wat|^2\times|\alpha|^2$ is the saturation parameter. The value
of $\alpha$ just deduced corresponds to the mean value of the field in
the very low saturation regime (Hechenblaikner \etal 1998). An iteration
is then started. These equations are valid for $s\ll 1$, so one can
put $<\!\sig_z\!>\approx-1/(1+s)$, we then deduce $<\!\sig\!>$ by
setting one more time $<\!\dot\sig\!>=0$ \eref{obsig} and use
\eref{fact} \emph{where the product is now treated as}
$<\!\aop\sig_z\!>=<\!\aop\!><\!\sig_z\!>=\alpha/(1+s)$. Those
procedures correctly normalize the populations, and we check that
\eref{obsz} is exactly cancelled if we use the values
($<\!\sig\!>,<\!\sig_z\!>$) just deduced and assume \eref{fact}
for $<\!\caop\sig\!>$. The iteration is stopped here without
treating \eref{oba} a second time, otherwise, as shown in the next
section, another state is obtained. The final equations are:
\ber
\label{RitschK} \alpha=\frac{\alpho}{1-\nu}\ ,\quad
<\!\sig_z\!>=-\frac{1}{1+s}\ ,\quad
<\!\sig\!>=\frac{g\alpha}{\wat}\frac{1}{1+s}\ .
\eer
\endnumparts
Those equations \emph{cannot} be deduced by \eref{ob}, where one would
set $s_{ob}=0$, because we are assuming  $s_{ob}\gg 1$ (an example is
illustrated in the next section).

 At this stage, three remarks are given. Firstly, as we treated equation
\eref{oba} only once, the value $<\!a\!>=-\alpha$ is the same in
\eref{Ritsch}\eref{RitschK} and it has been deduced by using the ground
state $|g\rangle$. As we assume low saturation, $|g\rangle$ is close to
the final state verifying \eref{RitschK}, but we distinguish it from the
latter because strictly speaking one has $<\!\sig_z\!>\neq-1$. The
ground state $|g\rangle$ plays the role of a \emph{referred state} that
was used to construct the final state. The status of the referred
states is strengthened below by the construction of other states.
The second point is that, the atomic mean values in \eref{RitschK}
correspond to optical Bloch equations where the atom is driven by
a classical field of amplitude $-\alpha$, hence the properties are
well known. Finally, the value of $\alpha$ determines entirely the
atomic quantities, thus all those steps can be reduced to two
steps: Firstly, one treats \eref{obsig}\eref{oba} by invoking
\eref{fact} and uses the referred state to deduce $\alpha=-<\!a\!>$.
From now on we write only the necessary equations:
\beq\eqalign{ \label{Heis}
<\!\dot\sig\!>=-i\wat(<\!\sig\!>-\,\frac{g}{\wat}<\!\aop\sig_z\!>)\\
<\!\dot\aop\!>=-i\wct(<\!\aop\!>\,+\,\alpho\,+
\frac{g}{\wct}<\!\sig\!>)\ .}
\eeq
Secondly, a Bloch equation description is assumed to deduce the atomic
mean values. In the next section, an optical Bloch Liouvillian with a
damped cavity mode is derived, its steady state $\rho_\alpha$ is a
\emph{factorized state}, it is known provided that one gives $\alpha$.
\subsection{General form for the steady states}
The method is based on a shift of the cavity mode by a
position-dependent complex number $\alpha\equiv\alpha(x)$:
\numparts
\ber
\label{ashift}
\aop\,\,=\,\,\cop-\alpha\ .
\eer
Such general shift is then used to describe equivalently the
Liouvillian \eref{Liou} with the Hamiltonian,
\ber
\fl\eqalign{
\label{Hamalpha} H = \wa\csig\sig\!-\!g\alpha\csig
\!-\!g\alpha^*\sig + \underbrace{g(\ccop\sig\!+\!\cop\csig)}
\,+\,\wc\ccop\cop +(E\!-\!\wct\alpha)\ccop +
(E\!-\!\wct\alpha)^*\cop\ ,}
\eer
with the dissipative terms ($\kap\Lc\,,\,\gam\Lsig$). Four components of
the cavity mode dissipative part $\La$ where already included in the
Hamiltonian through the imaginary part of the complex frequency
$\wct=\wc-i\kap$. If a given $\alpha$ is found such that the
contribution to the steady state $\rho_s$ from the coupling
$g(\ccop\sig\!+\!\cop\csig)$ cancels approximately the
contribution from the field driving term
$(E\!-\!\wct\alpha)\ccop+(E\!-\!\wct\alpha)^*\cop$, then $\rho_s$
can be approximated by the factorized steady state $\rho_\alpha$
of the Liouvillian $\Liou_\alpha$,
\ber
\eqalign{ \label{HamBloch}
\Liou_\alpha\rho_\alpha=0 ,\quad \Liou_\alpha\rho_\alpha =
 -i[ H_\alpha,\rho_\alpha]+\kap\Lc\rho_\alpha+\gam\Lsig\rho_\alpha=0\\
H_\alpha\ =\ \wc\ccop\cop + \wa\csig\sig  -g\alpha\csig
-g\alpha^*\sig\ .}
\eer
The system \eref{HamBloch} describes an atom driven by a
\emph{classical} field of amplitude $-\alpha$, i.e optical Bloch
equations, and a cavity mode representing a damped displaced field
($\dot c=-i\wct c$). The steady state of the cavity mode part of
\eref{HamBloch}, as much as the corresponding diffusion and friction
coefficients\footnote{For those dynamical quantities a similar
algebraical Liouvillian is used in sections
\ref{diffusion}-\ref{friction}}, can be described \emph{classically} by
the correspondance operator$\rightarrow$complex number,
$c\rightarrow -\alpha_c(t),\,\dot\alpha_c\!=\!-i\wct\alpha_c$. The
operatorial description of the field is nevertheless maintained for
convenience's sake because in \eref{HamBloch} the atom is quantized. The
steady state of $\Liou_\alpha$ is a factorized state and written
$\rho_\alpha=Q_\alpha\keto\brao$ where $\keto$ is the vacuum of
the displaced cavity mode, $c\keto\!=\!0$, it is a position
dependent coherent state of the annihilation operator $a$ with the
eigenvalue $-\!\alpha$, $a\keto\!=\!-\alpha(x)\keto$. $Q_\alpha$
represents the steady state of the atom and is entirely defined by
$\beta(x)$, which is in turn determined by $\alpha(x)$. One can
show that (see e.g Cohen-Tannoudji 1990):
\ber
\label{Bloch}
&&Q_\alpha=\big[\,\sig\csig+\beta\csig+
\beta^*\sig+|\beta|^2\,\big]\big/(1+s),\\
\label{beta}
&&\beta=\frac{g}{\wat}\alpha,\,\,s=2|\beta|^2\,,\qquad(\wat=\wa-i\gam)\
,
\eer
where $s\equiv s(x)$ is the atomic saturation parameter and
$1/(1\!+\!s)$ ensures a unit trace.
The expectation value of an operator $\mathcal{O}$ in the
steady state $\rho_\alpha$ is given by the trace
$\tr(\mathcal{O}\rho_\alpha)\equiv <\!\!\mathcal{O}\!\!>_\alpha$. Some
expectation values to be used in this paper are:
\ber
\label{expectation}
\eqalign{
<\caop\aop>_\alpha=|\alpha(x)|^2\ ,\quad
<\sig_z>_\alpha=-\frac{1}{1+s(x)}\\
<\sig>_\alpha=\frac{\beta(x)}{1+s(x)}\ ,\quad \ <F_d>_\alpha=
-2g(x)\frac{\wa}{\wa^2\!+\!\gam^2}\frac{|\alpha(x)|^2}{1+s(x)}\ ,
}
\eer
\endnumparts
where the force was calculated by \eref{force}. Since a state
$\rho_\alpha$ is entirely determined by a complex number
$\alpha=-<\!a\!>_\alpha$ we shall refer to $\alpha$ as a \emph{state}.

 The approximation $\rho_s\!\approx\!\rho_\alpha$ is valid under the
semiclassical approximation. It is in general satisfied if the
time scale that characterizes the internal evolution of the atom is well
separated from that of the cavity mode (bad or good cavity limits, large
atomic or cavity detuning). In practice, we estimate the validity of our
states by writing the dynamical equations
($\Liou_\alpha\rho_\alpha=0$):
\numparts
\ber \fl
\label{Heisapp1}
<\dot c>_\alpha=-i\wct(\alpho-\alpha+\frac{g}{\wct}<\!\sig\!>_\alpha)
\ ,\quad <\!\dot\sig\!>_\alpha=0\ ,\quad <\!\dot\sig_z\!>_\alpha=0\\
\fl\label{Heisapp2}
<\!\dot{\aop\sig_z}\!>_\alpha=-i\wct
(<\!\sig_z\!>_\alpha(\alpho\!-\!\alpha)\!-
\!\frac{g}{\wct}<\!\sig\!>_\alpha))\ ,
\eer
\endnumparts
and look under which parameter regime the equation for
$<\!\dot c\!>_\alpha $ (and $<\!\dot{\aop\sig_z}\!>_\alpha$) are totally
or partially cancelled. However, as higher order equations need to be
analysed, we compare our states $\rho_\alpha$ with the numerical
results.

 The determination of the different values for $\alpha$ is grounded on a
manipulation of the Hamiltonian \eref{Hamalpha} (and the dissipative
parts). Other values are deduced by applying a similar reasoning to the
Heisenberg equations of motion \eref{Heis}.
\subsection{An atom-cavity-mode cooperative picture: ($\nu,E$) and the
dressed states}
\label{ping-pong}
In this subsection, we manipulate the Liouvillian $\Liou$ and
show how $\nu$ \eref{NU} appears mathematically. We then show a
relationship between $\nu$ and $E$; and, finally, we relate
$\nu$ to the dressed states. The whole reasoning comes with an
intuitive picture, as well as indications on how states will be built
in the next sections.

$\nu$ can appear by two consecutive operators shifts.
The first shift corresponds to the choice $\alpha=E/\wct\equiv\alpho$
in\eref{Hamalpha}. The driving is entirely transferred to the atom,
$-\!g\alpho\csig\!-\!g\alpho^*\sig$, we cancel
this term by a second shift, on the atomic operators,
\ber
\label{Sig}
\sig\,\,=\,\,\Sig+\beta\ ,
\eer
with the convenient choice $\beta(x)=g\alpho/\wat\equiv\beta_0$ (see
\eref{beta}). The atomic driving is now exactly cancelled, and a new
term from the coupling Hamiltonian comes out. Such mode-driving term,
$\displaystyle{g\beta\ccop+g\beta^*\cop}$, is then written in terms of
$\nu$ \eref{NU} and $E$, to give:
\beq\eqalign{\fl \label{Ham2}
H = \wa\cSig\Sig + \wc\ccop\cop + g(\ccop\Sig +
\cop\cSig) + \nu E\ccop +\nu^*E^*\cop\ ,
\quad(\kap\Lc\,,\,\gam\LSig)\ .}
\eeq
In order to include the first \emph{reaction} of the atom on the cavity
mode, a third shift is operated, $c\rightarrow c-\nu\alpho$, to cancel
the mode driving term $\nu E$ in \eref{Ham2} (result not written).
Once that done, one returns from $\Sig$ to $\sig$, and checks that
these \emph{three local shifts} are equivalent to \emph{one global
shift}, starting from the initial Hamiltonian \eref{Hamalpha}, with
$\alpha=\alpho(1+\nu)$. As we shall see, this value of $\alpha$ gives
one simple and practical state. Another cycle will generate terms in
$\nu^2$ in \eref{Ham2}, and so on. By those three local shifts, the
strengh $E$ has been 'transported' and 'crossed' twice through the
coupling term in \eref{Hamalpha} (underbraced term). Actually, the
coupling term in \eref{Ham} plays the role of a pivot, it can be seen as
the net in a Ping-Pong game between the atom and the cavity mode, where
$E$ is the ball and $\nu$ is the dimensionless quantity that transports
the ball.

 An implicit relationship between $\nu$ and a finite value of $E$ is
addressed by the following two statements. Firstly, for $E\neq 0$, $\nu$
has a natural role, as shown in the scheme above or in the bistability
state equation \eref{ob}. Secondly, this is not the case when $E=0$,
unless for the trivial value $\nu=1$.
The second statement is explained by setting $E=0$ in \eref{Hamalpha}
and by shifting the atomic operators using \eref{Sig}. One then equates
each of the new driving terms,
$(\wat\beta-g\alpha)\csig$ and $(g\beta-\wct\alpha)\ccop$ (and hermitian
conjugates), to be zero\footnote{A second shift is necessary
to cancel the fictive driving created by the first shift.}. By
that, it is easy to show that the condition $\alpha\ne\!0$ implies
$\wat\wct\!=\!g^2$. But, this equality is impossible to satisfy unless
one has simultaneously $\kap\!=\!\gam\!=\!0$
($\Im(\nu)\!=\!0$) and $\wa\wc\!=\!g^2$
($\Re(\nu)\!=\!1$). From that brief analysis, we extract two
physical consequences. Firstly, by $E\!=\!0$ and $\kap\!=\!\gam\!=\!0$,
one has the dressed Hamiltonian \eref{Ham}, and, the condition
$\wa\wc\!=\!g^2$ means that one is tuning the lowest dressed
state $|-\rangle$ into resonance, in such case one has in fact
$\omega_{-}-\omega_{probe}=(\wa+\wc)/2-[4g^2+(\wa-\wc)^2]^{1/2}/2=0$.
Secondly, in \eref{Ham2}, as long as one decay rate only
($\kap$ or $\gam$) has a finite value, then $\nu\!\ne\!1$ and the atom
(or mode) can never return the same probe strength $E$ to the mode
(or atom) ($\nu E\ne E$).
 As $E$ and $\nu$ are somehow linked, we say that $\nu$ is also
the ball, it is \emph{the ball that moves}, and any
exponent of $\nu$ says how many exchanges took place. We call our method
the \emph{Ping-Pong}. Graphically one can see the local shifts as
a ping-pong exchange:
\numparts
\ber
\label{pingpongloc}
{}^{E\,\longrightarrow}\,{}^{\alpho}\!_\searrow
\nearrow{}^{\alpho\nu}\!_\searrow\nearrow{}^{\alpho\nu^2}\!_\searrow
\nearrow{}^{\alpho\nu^3}...\\
\quad\quad\quad\quad\!\!{}^{\beto}\quad\quad\quad\!\!\!{}^{\beto\nu}
\quad\quad\quad\!\!\!{}^{\beto\nu^2}\nonumber
\eer
or the global shifts with the reactions on the field only and where
the atom is represented by the hooked arrow,
\ber
\label{pingpongreac}
\alpho\hookrightarrow\alpho(1+\nu)\hookrightarrow
\alpho(1+\nu+\nu^2)\hookrightarrow ...\ ,
\eer
\endnumparts
the same last operation symmetric for the atom is of course possible by
substituting $\alpho$ for $\beto$. The starting in \eref{pingpongloc}
is explicitly shown on the field, this would be reversed if the atom
was driven first.
\section{The bounced states}
\label{bounce}
\subsection{First bounced state}
If one performs an infinite  iteration in \eref{pingpongreac}, a
known serie in $\nu$ (complex number) leads to the first bounced state
$\alpha=\alpha_{1b}$:
\numparts
\ber
\label{alphafrac}
\alpha_{1b}\,=\,\frac{\alpho}{1-\nu}\ .
\eer
Those infinite iterations could be avoided by making directly a shift of
$\sig$ in \eref{Hamalpha} and by cancelling the atom and mode driving
terms. The two conditions obtained
($E-\wct\alpha+g\beta\,=0,\,\wat\beta\,=\,g\alpha$) give back
\eref{alphafrac} if $\wct\wat\!\ne\!g^2$.

 That state is our example \eref{RitschK} in subsection \ref{Genesemi},
where the derivation has been opposed to the one that leads to the
bistability state equation \eref{ob}. We add some important
comparison with the derivation proposed by (Hechenblaikner \etal
1998). Their calculations purport that, under weak driving, the
Hilbert space can be restricted to the three states
$\{\left|g,0,bare\right>,\left|g,1,bare\right>,\left|e,0,bare\right>\}$,
where 'bare' means the Fock states in the basis of $\caop\aop$.
Thus, in the equation for $<\!\dot\sig\!>=\!0$ \eref{Heis}, the
equality $<\!\aop\sig_z\!\!>=\!-\!<\!\aop\!>$ is strictly
satisfied, and all expectation values factorize. By $<\!\dot\aop\!>=\!0$
\eref{Heis}, and by the replacement
$<\!\aop\!>\rightarrow\!-\alpha_{1b}$, we obtain \eref{alphafrac}.
Actually, their atomic steady state quantities differ from ours
\eref{alphafrac}\eref{HamBloch} owing to the presence of the saturation
parameter $s_{1b}\!=\!s\!\ne\!0$ in the denominators of
\eref{expectation}\footnote{The authors introduce the saturation
parameter, but only briefly in the bad cavity limit, and finally
linearize their equations by putting $s\!=\!0$ in the denominators in
order to obtain the friction and the diffusion coefficients.}.

 A part from that, firstly and by the operations above, the polarization
$<\!\sig\!>$ in the equation for $<\!\dot\sig\!>=\!0$ \eref{Heis} is
proportional to the field amplitude, $<\!\sig\!>\propto \alpha_{1b}$,
and hence \emph{it has been replaced by the variable $\alpha_{1b}$}.
Consequently, by looking at the equation for $<\!\dot\aop\!>=\!0$
\eref{Heis}, the cavity mode 'sees' itself because it is 'bounced' by
the atom. We call \eref{alphafrac} a \emph{bounced state}. Secondly,
such state is known to be valid for higher photon number, provided that
low saturation is ensured (Domokos \etal 2001). That means that a low
saturation regime can be chosen, but with high photon number,
$N_0\!>\!1$ and $|1\!-\!\nu|^2\!\le\!1$, in order to have the state
almost orthogonal to the subspace
$\{\left|g,0,bare\right>,\left|g,1,bare\right>,\left|e,0,bare\right>\}$.
Their derivation is restricted to weak driving, and hence we question
back the operation $<\!\aop\sig_z\!\!>=\!-\!<\!\aop\!>$, that is an
exact one in this subspace. The problem is absent in our description,
because, firstly, the state $\alpha_{1b}$ has its field part described
by a coherent state, which is, in the bare basis, centered around the
Fock state corresponding to a quantum number
$n\!\approx\!|\alpha_{1b}|^2$; secondly, the use of the referred state
$|g\rangle$ gives $<\!\aop\sig_z\!\!>=\!-\!<\!\aop\!>$ without any weak
driving assumption.

\begin{figure}
\centerline{\epsfig{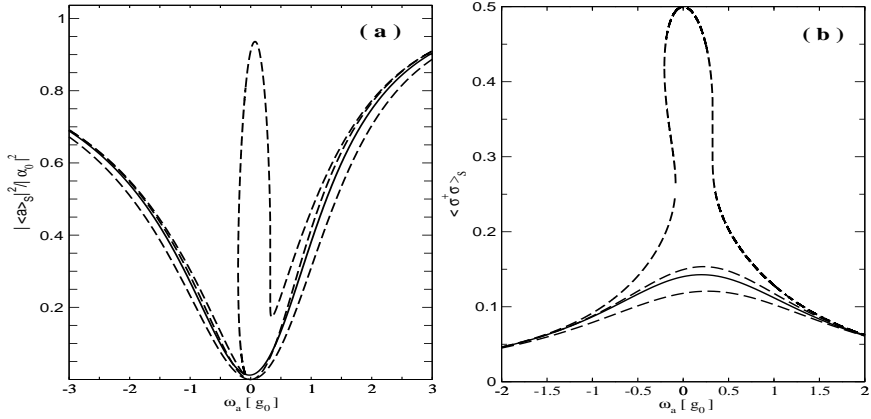}}
\caption{Comparing the bistability state equation with the exact result
and our approximations. Variation with respect to the atomic detuning
$\wa$. Figure(a), heterodyne transmission. From bottom to top at
$\wa=g_0$: state\eref{alphafrac}, exact state (solid line),
state\eref{bounce2}, bistability state equation\eref{ob}. Figure(b),
probability to be in the excited state. Same hierarchy. The parameters
are $(\gam\!=\!0.02,$$\kap\!=\!0.6,$$\wc\!=\!0.1,$
$g_0\!=\!1,$$N_0\!=\!0.37$).}
\label{LimitBistab}
\end{figure}

 In \fref{LimitBistab}, we plot the heterodyne transmission and the
probability of the excited state as a function of the atomic
detuning  $\wa$. The parameters are chosen in order to make the
bistability state equation  no longer valid ($s_{ob}\!\gg\!1$), whereas
our state (and the one derived below) reproduce reasonably well the
exact curve. For those parameters, one has
$<\!\aop\sig_z\!\!>_s\neq<\!\aop\!>_{ob}<\!\sig_z\!\!>_{ob}$ and
$<\!\aop\!>_{ob}<\!\sig_z\!\!>_{ob}\neq\!-\!<\!\aop\!>_{ob}$, whereas
the exact steady state verifies
$<\!\aop\sig_z\!\!>_s\approx\!-\!<\!\aop\!>_{1b}$. For larger detunings,
all curves meet, and the bistability state equation converges faster
than \eref{alphafrac}.

 Finally, we check the validity of that state. By \eref{Heisapp1},
$<\!\dot c\!>_\alpha\!=\!\varepsilon$ with
$\varepsilon=\!i\wct\alpho\nu[1\!-\!\nu]^{-1}\!\times\!s_{1b}/(1+s_{1b})
$. Thus, $\varepsilon\!\approx\!0$ for $(s_{1b}-0)\!\approx 0$ or
$\alpho\ll 1$ or $|\nu|\ll 1$, and, by generalization to other
Heisenberg equations, $\rho_s\!\approx\!\rho_\alpha$. Notice that
\eref{Heisapp2} is totally cancelled for the first bounced state, this
is also true for $<\!\dot{a\csig}\!>_\alpha\!=\!0$. Finally, bear in
mind the important condition $|\nu|\gg\!1$, or $|1\!-\!\nu|^2\gg\!1$
(for the photon number), that should ensure low photon number and low
saturation regimes; $\alpha_{1b}\!\rightarrow\!0$ (drop in
transmission), thus by \eref{beta}
$\beta_{1b}\!\rightarrow\!0\Rightarrow\!s_{1b}\!\rightarrow\!0$.
\subsection{Second bounced state}
This state is constructed by using the first one. As far as low
saturation is achieved, \eref{alphafrac} works for probabilities of
the excited state around $<\!\csig\sig\!>_{1b}\lesssim 0.1$. The
increase of the driving increases the difference $s_{1b}\!-0$, which we
interprete as a departure of that state (atomic part) from its own
referred state, i.e the ground state $|g\rangle$. The idea is, if
$s_{1b}\!-0$ is not too large, then one can iterate the procedure
by viewing the first bounced state as a referred one. Therefore, we
write in \eref{Heis} $<\!\aop\sig_z\!>=<\!\aop\!><\sig_z\!>_{1b}$ with
the population difference \eref{alphafrac}\eref{expectation}:
\beq
\label{sigzunsatbounced}
<\sig_z>_{1b}=-\frac{1}{1+s_{1b}}\ ,\qquad
s_{1b}(x)=\frac{2g^2(x)}{|\wat|^2}\frac{|\alpho|^2}{|1-\nu(x)|^2}\ ,
\eeq
and, as before, one deduces from \eref{Heis} the second state
$<\!\aop\!>\rightarrow-\alpha_{2b}$,
\beq
\label{bounce2}
\alpha_{2b}=\frac{\alpho}{1-\nu_{1b}} \,,\quad \mbox{where}\quad
\nu_{1b}(x)=\frac{\nu(x)}{1+s_{1b}(x)}\ .
\eeq
This state is still a bounced state, and we can see that it can be
deduced by a ping-pong scheme \eref{pingpongreac} with the simple
replacement $\nu\rightarrow\nu_{1b}$. The population difference of the
referred state $\alpha_{1b}$ is now transported. Contrary to the first
bounced state, the second one tends to $\alpho$ for increasing driving.
The parameter regime where such state is valid can be understood by
estimating as before \eref{Heisapp1}. By
\eref{bounce2}\eref{beta}\eref{expectation},
\beq
\label{sigbounce2}
S_{2b}\equiv <\!\sig\!>_{2b}\,\,=\,\,\frac{\beta_{2b}}{1+s_{2b}}
\,\,=\,\,\frac{g}{\wat}\frac{\alpha_{2b}}{1+s_{2b}}\ ,
\eeq
\endnumparts
with obvious notations, $<\!\dot c\!>_\alpha\!=\!\varepsilon$ with
$\varepsilon\propto\!\alpho(s_{2b}\!-\!s_{1b})$. Basically such state is
valid as far as the population of the excited state is close to that of
\eref{alphafrac}. In practice, it means that by keeping the atomic
detuning sufficiently large, the driving can be increased to reach
$<\!\csig\sig\!>_{2b}\lesssim 0.2$. The intermediate photon
number regime can be reached. In \fref{LimitBistab} this state is
presented and shows better agreement than \eref{alphafrac} because the
exact population of the excited state corresponds here to values around
$<\!\csig\sig\!>_{s}\approx\!0.15$.
\section{The polarized states}
\label{polarized}
An increase of the population of the excited state leads to the
appearance of the two level structure of the atom, as seen in the
equation $<\csig\sig>_s$$-<\csig>_s<\sig>_s=m(s)/2$, where
$m(s)\!=\!s^2/(1+s)^2$. For low saturation, $m(s)\!\sim\!s^2$, and
hence the two level structure of the atom can be represented by
the saturation parameter only. Bounced states have the property to
be defined only by the saturation parameter of the referred state, and
the increase of the population of the excited state transfers the
steady state $\rho_s$ from one bounced state to another. In such
regimes, the cavity mode amplitude is entirely returned by the
quasi-pointlike atom
($<\!\aop\sig_z\!>=<\!\aop\!><\sig_z\!>_{p}$), i.e $<\!\aop\!>$ is
a variable and $p$ is a fixed (referred) state. Else, the
polarization of an atom behaves as $<\!\sig\!>\sim\!s^{1/2}$ for
low saturation and as $<\!\sig\!>\sim\!s^{-1/2}$ for very high
one. Therefore, the trick is to say that a clear appearance of the
two level structure should be interpreted by treating the
polarization $<\!\sig\!>$ in the equation for $<\!\dot a\!>=0$
\eref{Heis} as one object that the field 'sees' as a whole. In
this case the polarization is the referred quantity.
\subsection{First polarized state}
The first polarized state corresponds to the first
hooked arrow in \eref{pingpongreac}:
\numparts
\ber
\label{alphasum}
\alpha_{1p}\equiv\alpho(1+\nu)\ .
\eer
The reactive term $\nu\alpho$ can be written $g/\wct\times S_{ref}$,
where $S_{ref}$ is a referred polarization of the atom because it
contains no information on the actual value of the mode, but rather,
it is incremented to the initial one $\alpho$
($S_{ref}\!=\!g/\wat\alpho$). The regimes where \eref{alphasum} is
valid is roughly estimated by calculating $<\!\!\sig\!\!>_{1p}$, to
find $<\!\dot c\!>_\alpha\!=\!\varepsilon$ where
$\varepsilon\!\propto\!-\nu\alpho[1\!-\!(1\!+\nu)/(1\!+\!s_{1p})]$.
Thus $\varepsilon\approx 0$ for $\nu\approx s_{1p}\,$. Such a condition
is a restriction but it can apply to high saturation regimes.

\begin{figure}
\centerline{\epsfig{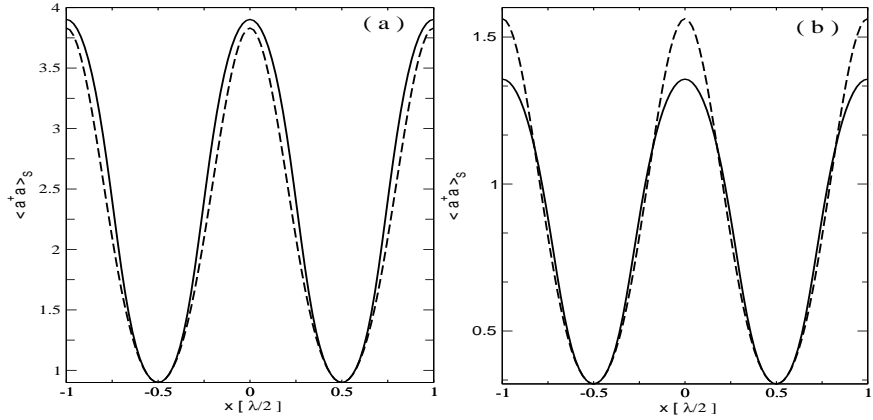}}
\caption{Mean photon number in the steady state as a function of the
axial position $x\,$. Figure(a), from (Pinkse \etal 2000)
$(\gam\!=\!0.187,$$\kap\!=\!0.087,$$\wa\!=\!2.8,$$\wc\!=\!0.31,$
$g_0\!=\!1,$$N_0\!=\!0.9$). Figure(b), from (Hood \etal 2000)
$(\gam\!=\!0.02,$$\kap\!=\!0.13,$$\wa\!=\!1.13,$$\wc\!=\!0.7,$$g_0\!=\!1
,$$N_0\!=\!0.32)$. The doted lines are $|\alpha_{1p}|^2\,$, the
analytical formula \eref{alphasum}. Solid lines are the exact numerical
calculations.}
\label{KimRempeNfot}
\end{figure}

 For the experiments of (Hood \etal 2000) and (Pinkse \etal 2000), the
probe is tuned to the lowest dressed state $|-\rangle$ that is, as
we said, the particular condition $\nu\approx 1$. As also
$s\!\approx\!1$, \eref{alphasum} should work to give
$<\!\caop\aop\!>_s\approx\!4N_0$ at an antinode. In
\fref{KimRempeNfot}, we plot the mean photon number in the steady
state as a function of the axial position of the atom. The agreement is
satisfactory, in particular the maximum and minimum that give good
agreement with the presented signals in both papers. For
(Hood \etal 2000), the transmission reaches about $1.2-1.3$, while for
(Pinkse \etal 2000) it goes around 4. State \eref{alphasum} is not an
approximation of \eref{alphafrac}, both states meet for
$\nu\!\rightarrow\!0$.
\subsection{Second and third polarized states}
More general states can be deduced by iterating the procedures above.
Bounced states are sensitive to resonances, such as $\nu\rightarrow 1$,
whereas the first polarized state \eref{alphasum} is not valid
for large values of $\nu$. The idea is to use both advantages.

 The second polarized state is calculated by reference to the
polarization \eref{sigbounce2} of the second bounced state
\eref{bounce2}. Consequently, this state is referred in a hierarchy to
the three states \eref{bounce2} \eref{alphafrac} and the ground state.
With the use of \eref{sigbounce2}, and then \eref{bounce2}, this gives
the second polarized state
($<\!\dot a\!>=0$ and $<\!a\!>\rightarrow -\alpha_{2p}$):
\ber\eqalign{
\label{standardbouncepolarization}
\alpha_{2p}\equiv\alpho+\frac{g}{\wct}S_{2b}
\,\,=\,\,\alpha_{2b}(1+\nu_q) \quad\nu_q=\nu_{2b}-\nu_{1b}\ ,}
\eer
where $\nu_{2b}=\nu/(1+s_{2b})$.
The last equation in \eref{standardbouncepolarization} shows that such
state is also deduced by a ping-pong exchange, with an input field being
now $\alpha_{2b}$ and a $\nu$ that transports a difference of population
difference (compare with \eref{alphasum}\eref{pingpongreac}). For
$s_{2b}\rightarrow s_{1b}$, the state reduces to $\alpha_{2b}$. In
general, \eref{standardbouncepolarization} is more flexible on the
detunings, it would work for probabilities of the excited state up to
$0.25\!-\!0.3$. For situations where the first bounced state
\eref{alphafrac} fails, the resulting polarized state
\eref{standardbouncepolarization} would smooth the results. That is the
case for the experiments we discuss.

 When such polarized state does not cancel resonant points, we can use
the first polarized state \eref{alphasum} as the original referred
state. To that end, \eref{alphasum} and \eref{expectation} are used to
form the saturation parameter $s_{1p}$ of the first polarized state; it
is then regarded as a referred state to deduce a third bounced state
$\alpha_{3b}$. We then calculate its polarization $S_{3b}$
\eref{expectation}, and view it as a referred state. One obtains thus
the third polarized state (and the third bounced state):
\ber
\label{polarized3}
&&\alpha_{3p}\equiv\alpho+\frac{g}{\wct}S_{3b}
\,\,=\,\,\alpha_{3b}(1+\nu_q) \quad\nu_q=\nu_{3b}-\nu_{1p}\\
\label{bounce3}
&&\alpha_{3b}=\frac{\alpho}{1-\nu_{1p}} \,,\quad
\nu_{1p}=\frac{\nu}{1+s_{1p}}, \quad
\big(s_{1p}=\frac{2g^2}{|\wat|^2}|\alpho|^2|1+\nu|^2\big)\ ,
\eer
\endnumparts
where $\nu_{3b}=\nu/(1+s_{3b})$. The third bounced state
\eref{bounce3}, that comes with the derivation, could work for the
low saturation regimes where the other bounced states do not
(typically $\nu\approx 1$). The third polarized state works well for the
experiments discussed here, in particular for the heterodyne signal.
However, as the atomic detuning is low, some difficulties are
encountered with the regime of (Hood \etal 2000). Some quantum effects
of the cavity mode are in fact present but we demonstrate below that
they are not needed for the dynamical quantities. Eventually, those
third states are like \eref{alphasum}, as they are not valid for large
values of $\nu$.
\begin{figure}
\centerline{\epsfig{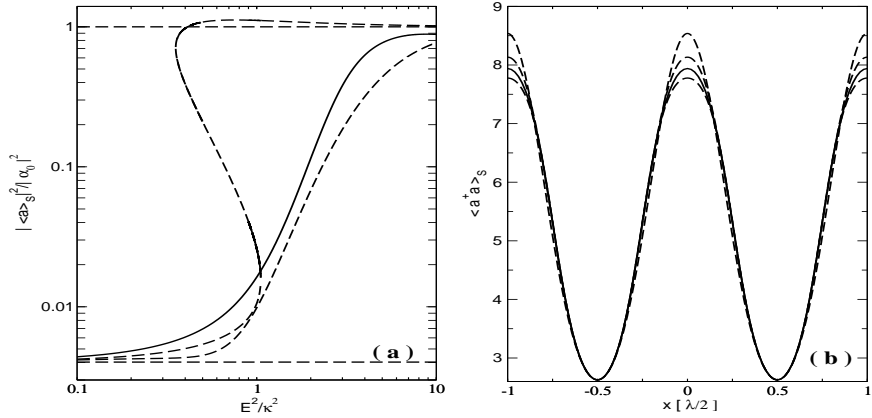}}
\caption{Figure(a), heterodyne signal as a function of the driving
strength. From (Hood \etal 1998),
$(\gam=0.02,$$\kap=0.33,$$\wa=\wc=0.166$,$g=g_0=1)$. From left to right,
at ordinate 0.1: bistability state equation\eref{ob}, exact state (solid
line), state\eref{standardbouncepolarization}. The lowest horizontal
line is the first bounced state \eref{alphafrac}. Figure(b), mean photon
number as a function of the axial position. Parameters,
$(\gam=0.187,\kap=0.094,\wa=6,\wc=1/6,g_0=1,N_0=2.63)$. From bottom to
top: state\eref{bounce3}, exact state (solid line), state
\eref{standardbouncepolarization}, state\eref{alphasum}. Notice the
difference between the maximum and minimum for the mean photon number
(intermediate regime).} \label{TrK6HD98} \end{figure}

 In \fref{TrK6HD98}(a), we plot the heterodyne transmission as a
function of the driving strength and compare it with three
approximations, among which \eref{standardbouncepolarization} and the
bistability state equation \eref{ob}. This plot is figure 5 of (Hood
\etal 1998), where the experimental data fit in well with the
theoretical quantum model \eref{Ham} and it constituted a part of the
experimental results that supported the conclusion on the quantum nature
of the field. But equation \eref{standardbouncepolarization} has a
classical-field interpretation. We mention, however, that other steady
state quantities are not well represented for these parameters.
In \fref{TrK6HD98}(b), we illustrate the intermediate photon number
regime, and with $\Re(\nu)\lesssim 1$. Such parameter regime could be
interesting experimentally, as it should offer a high signal to noise
ratio. The reason why these polarized states work efficiently comes from
the fact that we applied two different elementary procedures, the bounce
and the polarization. Therefore, such states are almost stable when the
parameters are varied. In counterpart, poorer results than with the
other states are to be expected though.

 In the next two sections are provided several expressions for the
diffusion and the friction coefficients. It is believed that the
arguments presented are easy to understand provided that one pays
particular attention to where stationary quantities are assumed, in
particular the different $\alpha$'s and $\beta$'s to be used. We repeat
that the results derived below are independent of any explicit
expression of the steady state quantities.
\section{Diffusion at zero velocity}
\label{diffusion}
In the papers of (Hood \etal 2000, Doherty \etal 2000), it is suggested
that the high suppression of the diffusion, when compared to the
semiclassical free space case, is a relevant signature of the quantum
nature of the field. As the mean photon number is around one, it was
concluded that one observed \emph{single atoms bound in orbit by
single photons} (Hood \etal 2000). The semiclassical free space
diffusion coefficient is calculated by assuming an equivalent standing
wave of strength $<\!a\!>_s$, driving the atom by $g\!<\!a\!>_s$. The
corresponding Liouvillian (Gordon and Ashkin 1980, Cohen-Tannoudji 1990)
is the atomic part in \eref{HamBloch}, with $-\alpha\rightarrow
<\!a\!>_s$. It leads to a diffusion about ten times (factor$\simeq7.5$)
bigger than the one obtained from the full calculus
\eref{Liou}\eref{Ham}. Consequently, the validity of \eref{HamBloch} is
questioned back, for the calculation of dynamical quantities, within the
cavity QED setting. One answer is addressed by pointing at a distinction
between dimensionless quantities and dynamical ones. Suppose a given
Liouvillian $\Liou$ with steady state $\rho_s$,\ $\Liou\rho_s=0$, with
eigenvectors and eigenvalues ($\rho^k\ ,E^k$),\ $\Liou\rho^k=E^k\rho^k$.
If it is possible to derive a scaled Liouvillian,
$\Liou_\lambda=\lambda\Liou$, then both Liouvillians have the same
steady state and eigenvectors, but the eigenvalues differ
$E^k_\lambda=\lambda E^k$. A dimensionless quantity, such as $<a>_s$, is
therefore unchanged, while a dimensioned one, as the diffusion, will
change according to its relation with the scaling procedure. More in
accordance with our case is to suppose a Liouvillian
$\Liou=\Liou_1+\Liou_2$ where ($\Liou_1,\Liou_2$) have the same steady
state $\rho_s\,$. If each sub-Liouvillian is scaled by different
factors, then $\Liou_\lambda=\lambda_1\Liou_1+\lambda_2\Liou_2$ still
has the same steady state $\rho_s$ as ($\Liou,\Liou_1,\Liou_2$) but the
eigenvectors will this time differ from those of $\Liou$. Dimensionless
quantities within the Hilbert space also change, except those that are
evaluated at the steady state. Starting from the full Liouvillian
\eref{Ham}, the approximated one \eref{HamBloch} offers a a good steady
state structure, but fails as far as the dynamics are concerned.
\subsection{Diffusion by eliminating the mode variables} A solution is
found by first noting that the atomic part of \eref{HamBloch} can be
written by using \eref{Sig}, ($\wa\cSig\Sig\,, \gam L_\Sig$) for the
Hamiltonian and the dissipative part respectively. The atomic steady
state $Q_\alpha$ \emph{verifies independently} $[\cSig\Sig,Q_\alpha]=0$,
and $L_\Sig Q_\alpha=0$. An effective atomic Liouvillian is then derived
by treating \eref{Heis} in the operatorial form (without noise terms),
by setting $\dot\aop=0$ in \eref{Heis} and by eliminating $a$ in the
equation for $\dot\sig$. As the equation for $\dot\sig_z$ is
unconsidered, then $\sig_z$ should be handled carefully. As long as
$\sig_z$ is alone, we keep it as an operator, and when it is in product
with $\sig$, it is 'frozen' by assuming that it takes a steady state
value $-1/(1+s$), given for example by some $\alpha$ and
\eref{expectation}. That reads: \numparts \ber
\dot\sig\,=\,-i\WAt\sig\,-ig\alpho\sig_z,
\quad\WAt=\wat(1-\frac{\nu}{1+s})\ ,\nonumber
\eer
which could be derived by the following Liouvillian,
\beq\eqalign{
\label{BlochScaled}
H_\alpha^{(at)}=\WA(x)\cSig\Sig,\quad
\GamA(x)L_\Sig,\quad\WAt=\WA-i\GamA\\
\WA(x)=\wa-\wc\frac{g^2(x)}{\wc^2+\kap^2}\cdot\frac{1}{1+s(x)}\\
\GamA(x)=\gam+\kap\frac{g^2(x)}{\wc^2+\kap^2}\cdot\frac{1}{1+s(x)}\ ,}
\eeq
where we already shifted $\sig$ by an amount
$\beta_h\equiv g\alpho/\WAt$, that is not related to $\alpha$ by
\eref{beta} but directly by a supplementary infinite ping-pong exchange
\eref{pingpongreac} with $\nu\rightarrow\nu/(1+s)$. As a rule, such a
difference will not affect the result (because $\beta_h\approx\beta$),
this supplementary bounce procedure can be seen as a way to seek
fluctuations around the steady state of \eref{BlochScaled}, which is
close to the one defined by $\alpha$. However, if the free space steady
state of \eref{HamBloch} is needed, the procedure will then be stopped
one step before, and hence $\sig_z$ will be frozen at some value "before
the steady state". Actually, this procedure is performed so as to avoid
non-linear equations. What is obtained is a new Liouvillian that is
scaled with respect to the free space case, where the dynamical
parameters ($\wa,\gam$) of the atom are modified by the inclusion of the
reaction of the field. As we saw, the dynamics is likely to change, but
with the same steady state. Finally, the dipole force is defined by:
\ber
\label{dipfd}
\Fcal\!=-(\nabla\!g) F_d, \quad F_d=-\alpha_d\csig\,-\alpha^*_d\!\sig\ ,
\eer
with a dimensionless atomic force operator $F_d$ approximated by
assuming a classical field characterized by $\alpha_d\ $, to be precised
below. The calculation is easily deduced from the free space case. The
diffusion tensor $D_{ij}$ is written:
\ber
\label{Dtensor}
D_{ij}=\big(\nabla g\big)_i\big(\nabla g\big)_j\times D\ ,\nonumber
\eer
where $D$ (hereafter called diffusion) is given in \ref{appdiff}. The
result is close to the free space expression, except in a global factor
slightly modified, plus a supplementary part that is in general
negligible. The free-space-like term $D_{at}(\alpha_d,s_h)\,$\eref{Dat}
can, in general, be approximated by $D_{at}(\alpha,s)\equiv
D_{at}(\alpha)\,$. It indicates that the force \eref{dipfd} is defined
by $\alpha_d=\alpha$, and the saturation parameter is displaced "one
step behind", thus returned to the steady state value. In
\fref{DKimRemp}(a) we plot $D_{at}(\alpha,s)$ for the experiment of
(Hood \etal 2000). The free space case reaches a value around $12\,$ at
the antinode (not shown). The suppression is well explained. By keeping
the term \eref{D0}, the diffusion gets closer to the exact one and even
closer if $s_h$ is left as it originally was (the diffusion will reach
the value $2$ at the antinode).

 For comparison's sake, it is also tempting to express the force
\eref{dipfd} in terms of $\alpho$, since everything is already included
in the modified atomic frequency and decay rate, and the atom is driven
by $-g\alpho$. In such a case, $D_{at}(\alpha_0,s)$ is lower than the
exact value. Actually, as \eref{BlochScaled} was used, we did not find
any proper argument to legitimate one value of $\alpha_d$ rather than
another in the expression of the force \eref{dipfd}. That is seen in
\fref{DKimRemp}(a) where different values of $\alpha_d=(\alpha,\alpho$)
approach the exact result differently. For these experiments, the
dominating term in $D_{at}(\alpha_d,s_h)$ is the known free space
diverging term in $s_h$: $\GamA^{-1}|\alpha_d|^2s_h^3[1+s_h^3]^{-1}$.
Here, formally, it does not diverge but reaches a constant value
$\GamA^{-1}|\alpha_d|^2$ (unless for $\alpha_d=\alpha$). The origin
of that divergence has been interpreted (Gordon and Ashkin 1980) in
terms of the dressed state approach in free space. However, the reason
for the suppression with regard to the experiment of (Hood \etal 2000)
is actually the value of $\GamA\approx10\gam$ at the antinode, thus
$D_{at}\approx D_{free}/10\ll D_{free}$.
\subsection{Diffusion by eliminating the atomic variables}
\begin{figure}
\centerline{\epsfig{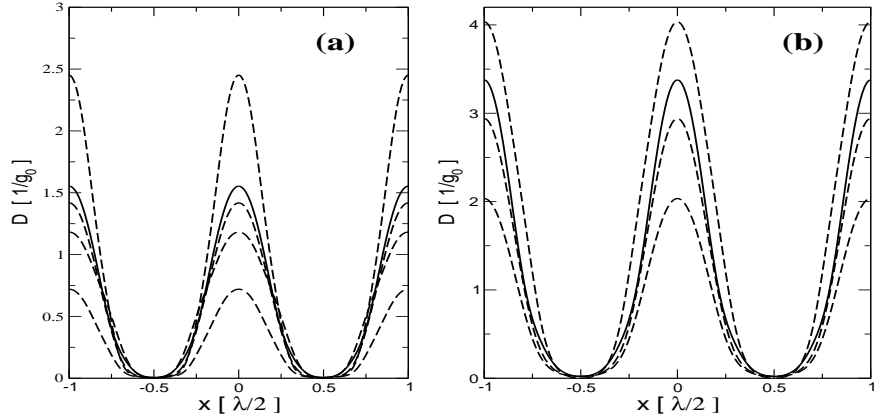}}
\caption{Diffusion coefficient $D$ as a function of the axial position.
Figure(a), parameters from (Hood \etal 2000). Figure(b), parameters from
(Pinkse \etal 2000). Same parameters as \fref{KimRempeNfot}. From bottom
to top: figure(a), $D_{at}(\alpho,s)$, $D_{cav}$\eref{Dcavsat},
$D_{field}\!+\!D_{at}(\alpho,s)$, $\mbox{exact (solid line)}$,
$D_{at}(\alpha,s)$. Figure(b), $D_{field}\!+\!D_{at}(\alpho,s)$,
$D_{field}\!+\!D_{at}(\alpha,s)$, $\mbox{exact (solid line)}$,
$D_{cav}$\eref{Dcavsat}. $\alpha$ is given by
\eref{standardbouncepolarization}.}
\label{DKimRemp}
\end{figure}

There is also a possibility to include all quantities in the field
variable, by setting $\dot\sig=0$ in \eref{Heis} and inserting it in the
equation for $\dot a$. Similar arguments to the previous case lead to:
\beq\eqalign{ \label{FieldScaled}
H_\alpha^{(field)}=\WC
(x)\ccop\cop\,\,,\quad \KC (x)\Lc\,,\ a=c-\alpha\\
\WC(x)=\wc-\wa\frac{g^2(x)}{\wa^2+\gam^2}\cdot\frac{1}{1+s(x)}\\
K_c(x)=\kap+\gam\frac{g^2(x)}{\wa^2+\gam^2}\cdot\frac{1}{1+s(x)}\ ,}
\eeq
where a simplification relating $\alpha$ to $s$ and $\beta$ has already
been made. The latter defines then the dipole force
$F_d=\beta\caop+\beta^*\aop$. This case is trivial (we scale out the
gradients):
\ber
\label{Dfield}
D_{field}=|\beta|^2\frac{K_c}{\WC^2+K_c^2}\ .
\eer
We plot in \fref{DKimRemp} $D\!=\!D_{field}\!+\!D_{at}(\alpha_0,s)$ for
both experiments, in \fref{DKimRemp}(b) we also show
$D\!=\!D_{field}\!+\!D_{at}(\alpha,s)$. State \eref{alphasum} can also
be used, the exact result will be approached differently.

 It should be clearly reminded that, in every case, Bloch equation-like
and a damped displaced cavity mode are the background assumption. A
picture is provided by imagining a swimmer swimming in a small pool, who
consequently generates waves which get reflected by the finite size of
the pool; that then changes the structure of the surface of the water to
finally make the swim more difficult to the swimmer by pushing him
according to the way he swims. In the end, the swimmer dissipates energy
($\GamA\gg \gam$) in trying to move and gets tired. In the free
space limit $\kap\rightarrow\infty$ , one has
($\WA,\GamA)\rightarrow(\wa,\gam$) and $K_c\sim\kap$, and hence
$D_{at}\rightarrow D_{free}$ and $D_{field}\rightarrow 0$.
\subsection{Diffusion by considering the atom-mode coupled system}
Those previous expressions for the diffusion supposed the field or the
atom as prescribed quantities. Other results are given where both
objects are regarded as "active". This is done by solving \eref{Heis}
approximately, the procedure is analogous to (Hechenblaikner \etal
1998). As the population difference $\sig_z$ is frozen, we consider two
cases that could be grouped into one in future works. The result for the
diffusion can be written $D=D_{cav}$:\\ with the "high saturation" case,
\ber
\label{Dcavsat}
D_{cav}=\frac{4\wa}{\wa^2\!+\!\gam^2\!}
\cdot|\alpha|^2\cdot
\frac{|\,\alpha\,|^2}{|\alpho|^2}\frac{\Im(\nu)}{(1+s)}\ ,
\eer
or the "low saturation" case,
\ber
\label{Dcavunsat}
D_{cav}=\frac{4\wa}{\wa^2\!+\!\gam^2\!}
\cdot\frac{|\alpha|^2}{(1+s)}\cdot
\frac{|\,\alpha\,|^2}{|\alpho|^2}\frac{\Im(\nu)}{(1+s)}\ .
\eer
\endnumparts
The difference is essentialy a factor $1/(1+s)$, whose origin is related
to the difference between $<\!\csig\sig\!>$ and $<\!\csig\!><\!\sig\!>$.
An additive term $\gam/|\wat|^2\times|\alpha|^2/(1+s)$ has been dropped,
it is a part of the free space expression and can be neglected within
our approximations. If one sets $s=0$, and explicits $\alpha$ by the
first bounced state expression \eref{alphafrac}, then one recovers the
expression derived by (Hechenblaikner \etal 1998). The authors
interpreted such term as \emph{cavity-induced}, we explicit here its
cooperative origin, indeed it is proportional to the imaginary part of
$\nu$. For $\kap\rightarrow\infty$, $\Im(\nu)\rightarrow 0$ and
$D_{cav}$ cancels. \Eref{Dcavsat} is plotted in \fref{DKimRemp}.
Eventually, the diffusion due to spontaneous emission is proportional
to the probability of being in the excited state, thus it is given by
\eref{expectation}.
\section{Friction at first order in velocity}
\label{friction}
The friction coefficient is derived from the same Liouvillians
\eref{BlochScaled}\eref{FieldScaled} as those for the diffusion, but we
use another definition for the force. Since the friction coefficient is
related to a velocity development, it is suitable to define here the
force by minus the gradient of the Hamiltonians
\eref{BlochScaled}\eref{FieldScaled}. That force being given, we develop
the expression to first order and relate its quantities to the spatial
derivatives of steady state mean values. The method is a standard
procedure, similar to (Gordon and Ashkin 1980, Hechenblaikner \etal 
1998). To perform the calculus, one suggestion is first to return from
the variables ($\Sig,c$) to ($\sig,a$) in \eref{BlochScaled} and
\eref{FieldScaled}, and then derive.  The velocity $\mathbf{v}$
dependent force $\digamma_v$ is given by the coefficient $G$ (hereafter
called friction):
\numparts
\beq
\label{frictionforce}
\digamma_v=- (\mathbf v\cdot\nabla g)\nabla g\,G\ .
\eeq
\subsection{Friction by eliminating the atomic variables}
The easiest case is the cavity-mode Liouvillian \eref{FieldScaled},
which leads to a force proportional to $\caop\aop$. By direct
derivation, one obtains an important expression:\footnote{Before the
derivation, put $(\WC-iK_c)\alpha\approx E=constant$}
\beq
\label{Gfield}\eqalign{
G_{field}=4|\alpha|^2\frac{\xi_c}{\WC^2+K_c^2}\Big (\
\frac{K_c\WC}{\WC^2\!+\!K_c^2}\,\xi_c
+\frac{1}{2}\frac{\WC^2\!-\!K_c^2}{\WC^2\!+\!K_c^2}\,\zeta_c\ \Big
)\\
\xi_c=2g\frac{\wa}{\wa^2+\gam^2}\frac{1}{1+s}F(s,\nu)\ ,\quad
\zeta_c=2g\frac{\gam}{\wa^2+\gam^2}\frac{1}{1+s}F(s,\nu)\ ,}
\eeq
where $s=2g^2|\alpha|^2/|\wat|^2\,$  and
($\xi_c(x),\zeta_c(x)\,,F(s,\nu)$) are discussed below. For our regimes,
it can generally be assumed $\zeta_c\!=\!0$. For
($s\!=\!0,F(s,\nu)\!=\!1$), $G_{field}$ reduces to the friction obtained
by (Hechenblaikner \etal 1998) in the good cavity limit.
\subsection{Friction by eliminating the field variables}
In the other limit \eref{BlochScaled}, the friction obtained is easily
interpreted as long as a free space term $G_{free}$ is clearly
recognized:
\beq
\label{Gat}
\eqalign{
G_{at}=\ G_{free}\
+\,\xi_a G_1+\xi_a^2 G_2
+\zeta_a (G_{1\zeta}\!+\!\xi_a G_{2\zeta})\\
\xi_a=2g\frac{\wc}{\wc^2+\kap^2}\frac{1}{1+s}F(s,\nu)\ ,\quad
\zeta_a=2g\frac{\kap}{\wc^2+\kap^2}\frac{1}{1+s}F(s,\nu)\ ,}
\eeq
\endnumparts
where ($G_{free},\,G_1,\,G_2,\,G_{1\zeta},\,G_{2\zeta}$)
are functions of $(s_h=2|\beta_h|^2,\GamA,\WA,g$) and given in
\ref{appfric}. Similarly to $G_{field}$, one can in general assume
$\zeta_a\!=\!0$ for $G_{at}$. For ($\xi_a\!=\!0,\zeta_a\!=\!0$),
the free space term is recovered, but it is defined by the modified
frequency $\WA$ and decay rate $\GamA$. For ($\alpho\rightarrow
0,s_h\rightarrow 0, s=0, F(s,\nu)\!=\!1$), $G_{at}/N_0$ \eref{GfreeG1G2}
gives back the expression derived by (Hechenblaikner \etal 1998) in the
bad cavity limit.

 In order totally to relate the friction coefficients
\eref{Gat}\eref{Gfield} to steady state quantities, one
still has to specify the function $F(s,\nu)$.
\subsection{Expressions for the function $F(s,\nu)$}
The dimensionless parameters $\xi_a\!\equiv\!-\dg\WA\,$,
$\xi_c\!\equiv\!-\dg\WC$ (and $\zeta_a,\zeta_c$) measure the
sensitivity of the atom and cavity frequencies to the variations of the
coupling. Some are oftentimes interpreted as trapping potentials
$U(x)$ acting on the atom, $\ \xi_c\equiv-\dg U(x)$. In the low
saturation limit, $\xi_c$ reduces to $2g\,\wa/(\wa^2+\gam^2)$, and can be
understood by analogy with a massive pointlike damped dipole oscillator
(Hechenblaikner \etal 1998). The inclusion of the saturation parameter,
through the real function $F(s,\nu)/(1\!+\!s)$, shows the departure from
which the two level structure appears. The function,
\ber
\label{Fsnu} F(s,\nu)=1-\frac{s}{1+s}\Re\big(\gdg\beta/\beta\big)\ ,
\eer
reduces to $1$ at zero saturation, and is related to the derivative of
$\beta$. If the steady state is provided by the ping-pong
schemes, then $F(s,\nu)$ contains the variation of the referred states.
However, in general, there is a possibility to provide several tractable
and efficient expressions for $F(s,\nu)$.
An approximation is given which is part of a generalized ping-pong
scheme to be dealt with elsewhere. One starts from a general bounced
state expression, $\alpha=\alpho/(1-\nu_p)$, and a polarized state
$\alpha=\alpho(1+\nu_p)$, with $\nu_p=\nu/(1+s_p)$, and where $s_p(x)$
is the saturation parameter relative to a referred state "p". The
derivatives $(\gdg\beta)/\beta$ become:
\ber\fl
\label{gdgbeta}
\gdg\beta/\beta\Big|_{bounce}=1+\frac{2\nu_p}{1\!-\!\nu_p}F(s_p,\nu)\
,\quad
\gdg\beta/\beta\Big|_{pol.}=1+\frac{2\nu_p}{1\!+\!\nu_p}F(s_p,\nu)\ ,
\eer
and hence $F(s,\nu)$ \eref{Fsnu} is related to the function
$F(s_p,\nu)$ \eref{gdgbeta} relative to the referred state "p". The next
step is to understand what kind of approximations are most efficient;
different possibilities are given:
\begin{figure}
\centerline{\epsfig{file=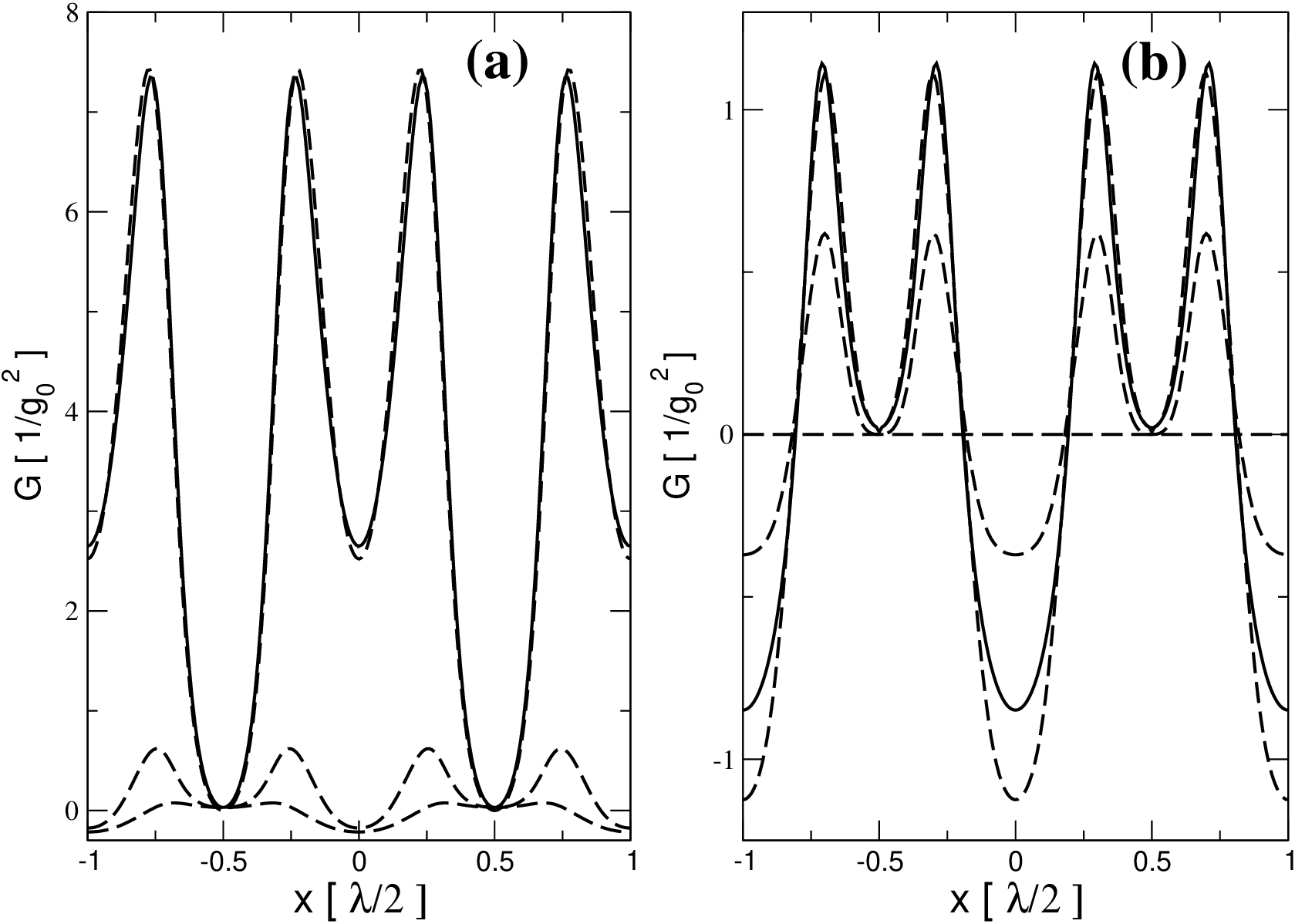,height=55mm,width=115mm}}
\caption{Friction coefficient $G$ as a
function of the axial position. Figure(a), (Pinkse \etal 2000). From
bottom to top: $G_{free}(s)$, $G_{at}(s)$, $ G_{field}$, exact $G$
(solid line). We used state \eref{standardbouncepolarization} and the
$\nu$ dependent version \eref{gdgbetasum} for the function $F(s,\nu)$.
Figure(b), (Hood \etal 2000). From bottom to top:
$G_{field}\!+\!G_{at}(s)$, exact $G$ (solid line), term $\xi_a
G_1\!+\!\xi_a^2 G_2\!+\!\zeta_a $$(G_{1\zeta}\!+\!\xi_a G_{2\zeta})$.
We used state \eref{standardbouncepolarization} and $F(s,\nu)$ is given
by \eref{gdgbeta1}. Notice, in figure(b), the positive contribution of
the translational friction term.}
\label{GKimRemp}
\end{figure}
\numparts
\ber
\label{gdgbeta1}
\gdg\beta/\beta\simeq 1\\
\label{gdgbetasum}
\gdg\beta/\beta\simeq 1+\frac{\nu}{1+s}
\rightarrow 1+\frac{1}{1+s}\\
\label{gdgbetafrac}
\gdg\beta/\beta\simeq 1+\frac{2\nu_\alpha}{1\!-\!\nu_\alpha}
\rightarrow 1-\frac{2}{1+s}\\
\label{gdgbetaob}
\gdg\beta/\beta\simeq
1+\frac{2\nu_\alpha}{1\!-\!\nu_\alpha}F(s,\nu)\ ,
\eer
\endnumparts
where $\nu_\alpha=\nu/(1+s)$. The referred state p has been eliminated.
The first case \eref{gdgbeta1} assumes a variation proportional to that
of a pure standing wave. The atom could be affected by the reaction
of the field but, when it moves, it sees a constant strength
($\beta\simeq g\times constant$). Such approximation is generally valid
for small values of $\nu$ or at high saturation (check with
\eref{gdgbeta}), as for (Hood \etal 2000). In that case, \eref{gdgbeta1}
gives $F(s,\nu)=1/(1+s)$. The two approximations in \eref{gdgbetasum}
are obtained with the polarized state expression, it is assumed that the
saturation parameter $s_p\approx s$ varies with $\nu$ as
$s\!+\!\nu\lesssim 1$, and $F(s_p,\nu)=1/(1+s)$ (pure standing wave).
Such approximation is more valid at intermediate saturation $\nu\approx
s\lesssim 1$. The arrow in equation \eref{gdgbetasum} means that, at
this step, $\nu=1$ can be assumed. Indeed, the functions
($\xi_a,\zeta_a,\xi_c,\zeta_c$) cancel for $g=0$ and hence the relevant
points are around the maximum of the coupling. In such a case,
\eref{Fsnu} simplifies to give $F(s,\nu)=1/(1+s)^2$, and the difference
of the exponent ($1+s$) with the function $F(s,\nu)$ derived from
\eref{gdgbeta1} shows again the limit between high and low saturation.
The third case \eref{gdgbetafrac} is aimed to work efficiently for large
values of $\nu$, it is derived from the general bounced state
expression. By the right arrow, we simplify the expression by taking the
limit $\nu_p\rightarrow\infty$, and a supplementary ratio $1/(1+s)$ has
been added from $F(s_p,\nu)$. The two arrowed limits just discussed have
an interest (a part simplification) because they are smoothed versions
of the expressions from which they derive, that might contain some
undesired resonance. Finally, \eref{gdgbetaob} represents the case close
to the optical bistability way-of-thinking, the value of $\beta$ is
returned, thus by \eref{Fsnu}, $F(s,\nu)$ can be deduced (not written).
Once these procedures made, the functions $F(s,\nu)$ now depend on the
saturation parameter of the steady state only, therefore one can simply
forget the origin of the states that lead to \eref{gdgbeta}. As final
simplifications, for the coefficients
($G_{free},G_1,G_2,G_{1\zeta},G_{2\zeta}$) \eref{GfreeG1G2}, one can as
usual operate $s_h\simeq s\,$ (as for the diffusion), thus \emph{all the
friction coefficients are now completely determined by providing
$\alpha$}.

\begin{figure}
\centerline{\epsfig{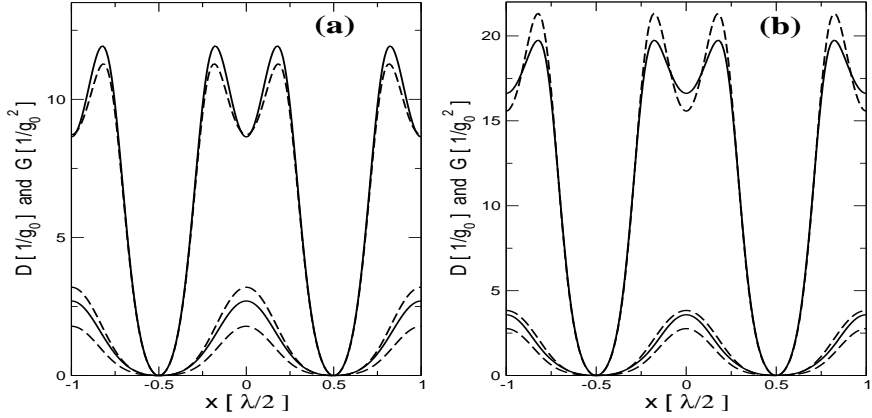}}
\caption{Diffusion and friction coefficients as a function of the axial
position. Figure(a), $(\gam=0.0236,\,\kap\!=\!0.13,\,\wa\!=\!6,\,
\wc\!=\!1/6,\, g_0\!=\!1,\,N_0\!=\!2.16)$. From bottom to top at $x=0$:
$D_{field}+D_{at}(s)$, exact $D$ (solid line),
$D_{cav}\eref{Dcavunsat}$, exact $G$ (solid line), $G_{field}$. For
$F(s,\nu)$, we used the $\nu$ dependent form of \eref{gdgbetafrac}.
Figure(b), same parameters as \fref{TrK6HD98}(b). From bottom to top at
$x=0$: $D_{field}+D_{at}(s)$, exact $D$ (solid line),
$D_{cav}$\eref{Dcavunsat}, $G_{field}$, exact $G$ (solid line). We used
the $\nu$ dependent form of \eref{gdgbetasum}. For both figures, we used
state \eref{alphasum}.} \label{GK6KR}
\end{figure}

 For the experiment of (Hood \etal 2000), the translated terms of
$G_{at}$ \eref{Gat} have an effect and tend to render the friction
positive. This is shown in \fref{GKimRemp} where we also plot the
friction for the experiment of (Pinkse \etal 2000). The friction
$G_{field}$ \eref{Gfield} is an excellent approximation for that
experiment. It shows that the quantum nature of the mode is again not
needed, thus responding to the questioning of (Doherty \etal 2000). A
basic result for this paper is plotted in \fref{GK6KR}, it shows a
situation of high positive friction and a diffusion similar to (Pinkse
\etal 2000). The temperature is about twice as low. Also notice that,
for \fref{GK6KR}(a), it is $G_{field}$ that works despite the fact that
$\gam<\kap$. This is a reminder that the bad/good cavity limits are only
rough limits, in \fref{GK6KR}(a) the opposite situation occurs. That is
caused by the other physical parameters, like the coupling $g$, which
are greater than $\kap$. Actually, this exact reversion can be explained
by extending the notion of bad/good cavity limits: \emph{($\GamA,K_c$)
are the relevant decay rates to determine which limit to use}. In
\fref{GK6KR}(a), we have in fact $\GamA > K_c$, thus the hierarchy is
reversed ($\gam<\kap$). Eventually, we present in \fref{DGNularge} a
cavity resonant case, with a high photon number $N_0\!=\!11$, high
friction, and a diffusion as low as (Hood \etal 2000). In that case, the
atom is trapped almost exactly half way between the node and the
antinode, thus it experiences a non zero friction and diffusion ($\nabla
g\neq 0$). Such parameters should be very efficient for trapping, with a
very clear drop at the output transmission.
\section{Remarks on optical bistability theory}
\label{sectionOB}
Our goal was first oriented towards cavity QED trapping mechanisms. It
was only when $\nu$ \eref{NU} was labelled and attributed some
importance, through the first ping-pong schemes
\eref{pingpongloc}\eref{pingpongreac}, that its interpretation became
needed. Its link to the cooperative parameter $C=g^2\big/2\kap\gam$
(Bonifacio and Lugiato 1978) is obvious and it is standard belief that
$\nu$ is a natural generalization of $C$. What is uncommon, however, is
that $\nu$ is never used in the optical bistablity theory (Lugiato and
Narducci 1990, Abraham and Smith 1982), as far as we are concerned. We
make essentially two remarks, showing that, maybe, one could rethink
optical bistability with $\nu$ and $C$ in hand. There is a standard
mean-field optical bistability state equation, which is commonly written
in the form (Lugiato and Narducci 1990):
\begin{figure}
\centerline{\epsfig{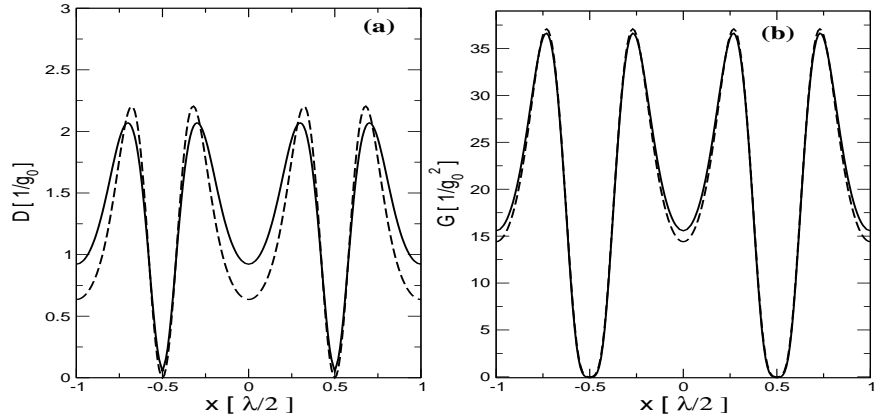}}
\caption{Diffusion and friction coefficients as a function of the axial
position. Cavity resonant case.
$(\gam=0.18,\,\kap\!=\!0.09,\,\wa\!=\!-5,\, \wc\!=\!0,\,
g_0\!=\!1,\,N_0\!=\!11)$. From bottom to top at $x=0$: Figure(a),
$D_{cav}$\eref{Dcavunsat}, exact $D$ (solid line). Figure(b),
$G_{field}$, exact $G$ (solid line). For $F(s,\nu)$, we used the $\nu$
dependent form of \eref{gdgbetafrac}, with state \eref{bounce2}. The
atom should be trapped at points $x_M$ such that $g^2(x_M)=-\wa\kap$,
with a temperature $D/G\approx 2/35g_0\simeq \kap$.}
\label{DGNularge}
\end{figure}
\ber
x=y\{1+i\theta +2C(1-i\delta)/(1+\delta^2+|x|^2)\}^{-1}\ ,
\eer
where $x,y\,$ are the scaled output and input field amplitudes
respectively. That equation depends on the scaled detunings
$\theta=\wc/\kap, \delta=\wa/\gam$ and the cooperative parameter $C$.
Therefore, the space of parameters in which one works is
the $(C,\delta,\theta,y)$ space; however, as one parameter is
usually a variable, like for example the input amplitude $y$,
bistability conditions are drawn in the cubic space $(C,\delta,\theta)$.
We showed that the state equation above is actually the bounced state
\eref{ob}, scaled differently with the schematical correspondences
$y\rightarrow\alpho,\,x\rightarrow\alpha_{ob}$, and with a saturation
parameter related to $|x|^2$. The referred state is here the
state itself. Consequently, the space is now reduced to $(\nu,\alpho)$
or ($\nu$) alone if $\alpho$ is a variable. What was to be expected
does happen; we show two typical inequalities that are conditional to
optical bistability (Agrawal and Carmichael 1979, Drummond and Walls
1981): \numparts
\ber
4\big[\,\delta\theta+C-1\,\big]^3 \geq
27C(\delta^2+1)(\theta^2+1)\,,\quad 2C\geq\delta\theta-1\ .
\eer
Those two conditions could be re-written,
\ber
\big[\,|\nu|^2\,+\,2\Re(\nu)\,\big]^3 \geq
27|\nu|^4\,,\quad|\nu|^2\geq\Re(\nu)\ .
\eer
\endnumparts
That is a strong result. Those bistability conditions depend essentially
on the value of $\nu$, thus we gain one dimension since there is a
possibility to draw the conditions in the $\nu\!-\!$complex plane.
Notice incidentally that the conditions above relate the phase of
$\nu\,$ to its modulus. How far such simplifications go should be
checked in order to determine the exact status of $\nu$ and $C$.
\section{Conclusions}
\label{conclusion}
If there was one conclusion to draw it would be the reminder that, even
if quantum effects happen to be in the cavity QED setting, the
confinement of the electromagnetic field within boundaries gets along
with a strong classical behaviour. That can be interpreted as, due to
the boundaries, even one photon on average would meet several times the
atom, the latter sees a lot of photons, thus most likely a classical
field. We showed that, even if one tunes the external laser
quasi-resonantly with a dressed state, the coherent driving has a
continuous component that almost washes away those quantum states. That
is true for the steady state and for the diffusion and friction
coefficients. We provided a self-consistent method that recognizes a new
physical parameter and that allows one to extend the semiclassical limit
out of both the bistability state equation domain and the low saturation
regime. We derived classes of steady states that incremently took into
account the quantum nature of the atom, while the field remained
classical. Besides, we gave diffusion and friction coefficients that do
not depend on any particular form of the steady state. We pointed out
that the usual bad/good cavity limits can be reformulated in the strong
coupling regime by the statement: The modified decay rates ($\GamA,K_c$)
become the relevant decay rates to determine which limit to use.

 The steady states are functions of the \emph{referred states}, which in
general are steady states but for other parameter regimes. The steady
states belong to two distinct families, the \emph{bounced states}, and
the \emph{polarized states}. They respectively refer to the
\emph{saturation parameter} and the \emph{polarization} of the
lastly referred state. The parameter $\nu$ is a \emph{step} that allows
one to move from one state to another, it also represents the quantity
that sends back the driving strength to the mode (or the atom) from $E$
to $\nu E$. Thus, the method is called the \emph{Ping-Pong} since the
coupling $g(\caop\sig+\aop\csig)$ may represent the net of a ping-pong
game between the atom and the mode, where ($E,\nu$) respectively
represent the ball and the transporter. Although $\nu$ does not depend
explicitly on $E$, it however relies on a finite value of the driving
strength $E\ne 0$. Such relationship between $\nu$ and $E$ has as its
cornerstone the lowest dressed state. The condition $E=0$ gets along
with $\nu=1$, the latter equality meaning that the lowest dressed state
$|-\rangle$ is tuned into resonance. In parallel, $\nu$ seems to have a
deeper meaning than the cooperativity parameter $C$. We showed that
known inequalities that are conditional to optical bistability are
actually functions of the complex parameter $\nu$ \emph{only}, whereas
they are usually written in terms of ($C,\delta,\theta$).

	We want to notice that the pictures which appeared in this paper,
whether they are the ping-pong or the swimmer, are descriptive of the
\emph{structure} of the system. They are believed to be complementary to
the sisyphus effect which is more a \emph{dynamical} picture of the atom
climbing up and down potential hills. A structure effect has been
explicited by showing that the reduction of the diffusion for the
experiment of (Hood \etal 2000) can be explained by a scaling argument
on the dynamical variables due to the interactivity between the atom and
the mode, and so such a reduction does not seem to be a mystery anymore.
Similarly, the method is able to recover accurately the experiment of
(Pinkse \etal 2000). We also derived a state, with a classical
assumption for the field, which reproduces reasonably well a heterodyne
transmission measured by (Hood \etal 1998) whereas the bistability state
equation is no longer valid. The understanding of the structure of the
system should be deeply thought in future works because it lays the
stress on settled (permanent) quantities, that are fixed as soon as
dimensionless parameters are. We shall show elsewhere that there are an
infinite number of steady states exhibiting fractal structures, which
could hopefully account for instabilities in cavities. The ping-pong
method is extendable to linear systems having a coupling term of the
form discussed in this paper and thus a wide range of electrodynamical
Liouvillians could be addressed. The classical assumption for the field
is relevant for this paper only, the ping-pong method can be applied to
quantum mode regimes. In our opinion, should future experiments in
cavities be done to trap an atom by a quantum field, then one should
test first a clear signature of the quantum origin of the mode by making
for example a heterodyne and photodetection measurements. If those
signals are clearly separated, the atom will possibly be trapped by a
quantum field. None of those experiments clearly fulfilled such kind of
conditions.

\ack
Entangled thanks to Paolo Tombesi, David Vitali and G\'erard
Massacrier for interesting discussions on physics and for the
preparation of this paper. Many thanks to the singing people with whom
I shared a lot of sympathy in this peaceful part of Italy. This work is
supported by INFM.
\appendix \section{Results for the diffusion and friction coefficients}
\subsection{Diffusion coefficient, atomic part}
\label{appdiff}
Solution for the diffusion coefficient that is a
generalization of (Gordon and Ashkin 1980), for the standing wave case,
with the Liouvillian given by:
\ber
\label{HamBlochMedAtApp}
H = \WA(x)\csig\sig -g\alpha_h(x)\csig -g\alpha_h^*(x)\sig \,\,,\quad
\GamA(x)\Lsig\ ,
\eer
where there is no need to specify the exact form of
the frequency $\WA(x)$ and the decay rate $\GamA(x)$. The force is
given by \eref{dipfd}. The procedure is identical to (Gordon and Ashkin
1980), at the exception that the distinction between $\alpha_h$ and
$\alpha_d$ leads to two suplementary terms here grouped in
$D_0$. The result is for the diffusion \eref{Dtensor},
$D=D_{at}+\Re(D_0)$ :
\ber\fl\eqalign{\label{D0}
D_0=\frac{1}{\GamA}\frac{g^2}{\WA^2+\GamA^2} \frac{\Delta_\alpha}{1+s_h}
\Big(\,1-\frac{1}{(1+s_h)^2} \big(\,1+\frac{4\GamA^2}{(\WA\!-i\GamA \!
)^ 2 } +2s_h\frac{\WA}{\WA\!-i\GamA\!}\big)\Big)\ ,}
\eer
where
$\Delta_\alpha=\,\alpha_h^2\alpha_d^{*2}- |\alpha_h|^2|\alpha_d|^2$. The
free space like term is given by:
\ber\fl\eqalign{ \label{Dat}
D_{at}(\alpha_d,s_h)=\frac{\GamA}{\WA^2\!+\!\GamA^2\!}
\frac{|\alpha_d|^2}{(1+s_h)^3} \Big(1+
\big(\frac{4\GamA^2}{\WA^2\!+\!\GamA^2\!}-1\big)s_h\,
+3s_h^2\,+\frac{\WA^2\!+\!\GamA^2\!}{\GamA^2}s_h^3\Big)\ ,}
\eer
where
$s_h=2|\beta_h|^2$ with $\beta_h=g\alpha_h/\WAt,\ \WAt=\WA-i\GamA$.
\eref{Dat} is sufficient with $s_h=s\,$ and $\alpha_d=(\alpha\
\mbox{or}\ \alpho)$.
\subsection{Friction coefficient, atomic part}
\label{appfric}
The friction is calculated by starting with \eref{BlochScaled}, or
directly by \eref{HamBlochMedAtApp} with $\alpha_h=\alpho$. The force
operator defined by minus the gradient of the Hamiltonian is developped
to order one in velocity:
\ber
&&G_{at}=-\alpho\csig_v-\alpho^*\sig_v-\,\xi_a(x)(\csig\sig)_v\ ,
\eer
and is expressed in terms of the derivatives of the steady state values
of ($\sig,\,\csig,\,\csig\sig$).
\ber
&&(\csig\sig)_v=\frac{1}{4\GamA}\frac{1}{1+s_h}\Big(
-\dg\!\!<\sig_z>_s+4\Re\Big[\alpho^*g\dg\!\!<\sig>_s\!\!\big/\WAt\Big]
\Big)\nonumber\\
&&\sig_v=-2\beta_h(\csig\sig)_v+i\dg\!\!<\sig>_s\!\!\big/\WAt\
.\nonumber
\eer
Use $\beta_h=g\alpho/\WAt$, and its derivative
$\gdg\beta_h=\beta_h(1+g(\xi_a+i\zeta_a)/\WAt)$, where $\xi_a,\zeta_a$
are related to $F(s,\nu)$ in \eref{Gat}. The expression for the
coefficient $G_{at}$ \eref{Gat} finally reads:
\beq\fl\eqalign{
\label{GfreeG1G2}
G_{free}(s_h)=\frac{1}{\WA^2\!+\!\GamA^2}
\frac{2|\alpho|^2}{(1\!+\!s_h)^3}\frac{\WA}{\GamA} \Big\{-s_h^2 +
2(1\!-\!s_h)\frac{\GamA^2}{\WA^2\!+\!\GamA^2}\Big\}\\
G_1(s_h)=\frac{1}{\WA^2\!+\!\GamA^2}
\frac{2|\alpho|^2}{(1\!+\!s_h)^3}\frac{g}{\GamA} \Big\{\
-s_h^2+\frac{1}{2}s_h+ (4\!-\!3s_h)\frac{\GamA^2}{\WA^2\!+\!\GamA^2}
-\frac{4\GamA^4}{(\WA^2\!+\!\GamA^2)^2}\Big\}\\
G_2(s_h)=\frac{1}{\WA^2\!+\!\GamA^2}
\frac{s_h}{(1\!+\!s_h)^3}\frac{\WA}{\GamA} \Big\{\
\frac{1}{2}s_h+\frac{2\GamA^2}{\WA^2\!+\!\GamA^2}\Big\}\\
G_{1\zeta}(s_h)=\frac{1}{\WA^2\!+\!\GamA^2}\frac{2|\alpho|^2}{(1\!+\!s_h
)^3} \frac{g\,\WA}{\WA^2\!+\!\GamA^2}
\Big\{1\!-\!\frac{4\GamA^2}{\WA^2\!+\!\GamA^2}\Big\}\\
G_{2\zeta}(s_h)=\frac{1}{\WA^2\!+\!\GamA^2}
\frac{s_h}{(1\!+\!s_h)^3}\Big\{
1\!+\!\frac{1}{2}s_h\!-\!\frac{2\GamA^2}{\WA^2\!+\!\GamA^2}\Big\}\ .\\ }
\eeq
The analytical expressions for $G_{at},G_{field}$ are explicited by
providing a particular formula for $F(s,\nu)$. The form of the functions
$\WA (x),\GamA (x)$ in \eref{BlochScaled} only affect the
functions $\xi_a (x),\,\zeta_a (x)$.
\section*{References}
\begin{harvard}
\item[] Abraham E and Smith S D
1982 {\it\RPP} {\bf 45} 815
\item[] Agrawal G P and Carmichael H J 1979
{\it\PR A} {\bf 19} 2074
\item[] Bonifacio R and Lugiato L A 1975
{\it\PR A} {\bf 11} 1507
\item[] \dash  1978 {\it\PR A} {\bf 18} 1129
\item[] Cohen-Tannoudji C,
1990 {\it Fundamental Systems in Quantum Optics}, Les Houches Session
LIII, ed Dalibard J \etal (North-Holland, eds Elsevier Science,1992)
\item[] Doherty A C, Parkins A S, Tan S M and Walls D F 1997
{\it \PR A} {\bf 56} 833
\item[] Doherty A C, Lynn T W, C J Hood and Kimble H J 2000 {\it \PR A}
{\bf 63} 013401
\item[] Domokos P, Horak P, Ritsch H 2001 {\it
J. Phys. B: At. Mol. Opt. Phys.} {\bf 34} 187
\item[] Drummond P D and Walls D F 1981 {\it \PR A}
{\bf 23} 2563
\item[] Gordon J P and Ashkin A 1980 {\it \PR A} {\bf 21} 1606
\item[]Hechenblaikner G, Gangl M, Horak P and Ritsch H 1998 {\it Phys.
Rev. A} {\bf 58} 3030
\item[] Hood C J, Chapman M S, Lynn T W and Kimble H J 1998
{\it \PRL} {\bf 80} 4157
\item[] Hood C J, Lynn T W, Doherty A C, Parkins A S and Kimble H J 2000
{\it Science} {\bf 287} 1447
\item[] Horak P, Hechenblaikner G, Gheri K, Stecher H and Ritsch H
1997 {\it \PRL} {\bf 79} 4974
\item[] Kimble H J lecture, 1994 {\it Cavity
Quantum Electrodynamics} (edited by Berman P) (Advances in Atomic,
Molecular, and Optical Physics, Supplement 2)(Academic, New York)
\item[] Lugiato L A and Narducci L M, 1990 {\it Fundamental Systems in
Quantum Optics}, Les Houches Session LIII, ed Dalibard J \etal
(North-Holland, eds Elsevier Science,1992)
\item[] Pinkse P W H, Fisher T, Maunz P and Rempe
G 2000 {\it Nature} {\bf 404} 365
\item[] Vuleti\'c V and Chu S 2000 {\it \PRL} {\bf 84} 3787
\end{harvard}
\end{document}